\documentclass[useAMS,usenatbib]{mn2e}

\usepackage{epsfig}
\usepackage{amsmath}
\usepackage{amssymb}
\usepackage{graphicx}
\usepackage{subfig}

\def\apj{ApJ}
\def\apjl{ApJL}
\def\mnras{MNRAS}
\def\pasp{PASP}

\def\aapr{AAPR}
\def\araa{ARAA}
\def\aap{A\&A}
\def\aj{AJ}
\def\apjs{ApJS}

\def\nat{Nature}

\def\nar{NAR}

\def\um{\mathrel{\rm \mu m}}
\def\ergss{\mathrel{\rm erg \; s^{-1}}}
\def\lx{\mathrel{\rm L_{2-8keV}}}
\def\lagn{\mathrel{\rm L_{AGN}}}
\def\lir{\mathrel{\rm L_{IR,SF}}}

\def\mlir{\mathrel{\rm \langle L_{IR,SF} \rangle}}
\def\mlx{\mathrel{\rm \langle L_{2-8keV} \rangle}}
\def\mlagn{\mathrel{\rm \langle L_{AGN} \rangle}}

\title[The SFRs of X-ray AGN]{A remarkably flat relationship between the average star
  formation rate and AGN luminosity for distant X-ray AGN}

\author[F.\ Stanley, et al.]
{ \parbox[h]{\textwidth}{ 
F.\ Stanley,$^{\! 1\, *}$
C.\ M.\ Harrison,$^{\! 1}$
D.\ M.\ Alexander,$^{\! 1}$
A.\ M.\ Swinbank,$^{\! 1}$
J.\ A.\ Aird,$^{\! 1,2}$
A.\ Del~Moro,$^{\! 1}$
R.\ C.\ Hickox,$^{\! 3}$
and J.\ R.\ Mullaney$^{\! 4}$ 
}
\vspace*{6pt} \\
$^1$Centre for Extragalactic Astronomy, Department of Physics, Durham University, South Road, Durham, DH1 3LE, UK \\
$^2$Institute of Astronomy, University of Cambridge, Madingley Road, Cambridge, CB3 0HA, UK \\
$^3$Department of Physics and Astronomy, Dartmouth College, 6127 Wilder Laboratory, Hanover, NH 03755, USA \\
$^4$Department of Physics \& Astronomy, University of Sheffield, Hounsfield Road, Sheffield, S3 7RH, UK\\
$^*$Email: flrstanley@gmail.com \\
}

\begin{document}
\maketitle

\begin{abstract}

In this study we investigate the relationship between the star formation rate (SFR) and AGN luminosity ($\lagn$) 
for $\sim$2000 X-ray detected AGN. The AGN span over three orders of magnitude in X-ray luminosity ($10^{42} < \lx < 10^{45.5} \ergss$) 
and are in the redshift range $z =$ 0.2 -- 2.5. Using infrared (IR) photometry (8 -- 500\,$\um$), including deblended 
\textit{Spitzer} and \textit{Herschel}  images
and taking into account photometric upper limits, we decompose the IR spectral energy distributions 
into AGN and star formation components. Using the IR luminosities due to star formation, we investigate the average 
SFRs as a function of redshift and AGN luminosity. In agreement with previous studies, we find a strong evolution 
of the average SFR with redshift, tracking the observed evolution of the overall star forming galaxy 
population. However, we find that the relationship between the average SFR and AGN luminosity is broadly flat at all 
redshifts and across all the AGN luminosities investigated; in comparison to previous studies, we find less scatter amongst the
average SFRs across the wide range of AGN luminosities investigated. By comparing to empirical models, we argue that the observed 
flat relationship is due to short timescale variations in AGN luminosity,
driven by changes in the mass accretion rate, which wash out any underlying 
correlations between SFR and $\lagn$. Furthermore, we show that the exact form of the predicted relationship between 
SFR and AGN luminosity (and it's normalisation) is highly sensitive to the assumed intrinsic Eddington ratio distribution. 

\end{abstract}


%%%%%%%%%%%%%%%%%%%%%%%%%%%%%%%%%%%%%%%%%%%%%%%%%%%%%
\section{Introduction}
%Key problem: AGN -- SF connection
%   -results serving as broad constraints 
%   -dependence of both on cold gas supply -->  not surprising that we find a broad connection 
%   -no direct evidence of this connection <-- where does our work fit in.
One of the key outstanding problems in studies of galaxy evolution is understanding the
connection between active galactic nuclei (AGN) and star formation.
Both AGN activity and star formation are predominately dependent on the 
availability of a cold gas supply from the
galaxy, as it is the fuel of both processes, and therefore a first order 
connection between these two processes may be expected. 
However the scales of AGN activity and star formation 
are very different, which has lead to suggestions that 
any tight connection between them must be due to one process regulating
the other (see \citealt{Alexander12}, \citealt{Fabian12}, and \citealt{Kormendy13} for recent reviews).

There are several pieces of empirical evidence for at least a broad
connection between AGN activity and star formation.
For example, the tight correlation observed 
between the mass of the super-massive black hole 
(SMBH) and the galaxy spheroid for galaxies
 in the local universe (e.g., \citealt{Kormendy95},
\citealt{Magorrian98}), serves as archaeological 
evidence of a connection between the growth of 
the SMBH (through mass accretion,
where it becomes visible as AGN activity), 
and the growth of the galaxy (through star formation). 
Additionally, observations of 
AGN have found that the volume average of the SMBH mass 
accretion rate tracks that of the star formation rate (SFR), within $\sim$3--4 orders
of magnitude, up to redshifts of $z \sim$2
(e.g., \citealt{Heckman04}; \citealt{Merloni04}; \citealt{Aird10}) 
suggesting a co-evolution of AGN and star forming activity.  
Despite how significant these
results may appear, they only provide indirect evidence
for a relationship between AGN activity and star formation and 
cannot place strong constraints on the form of the relationship. 

% 
%Tools : Why X-ray and FIR (Majority of SMBH growth occurred at z ~1
%etc)
To acquire more direct evidence on the form of the relationship between AGN
activity and star formation requires sensitive measurements 
of the AGN and star forming luminosities of individual galaxies.
X-ray and far--infrared (FIR; $\lambda = 30 - 500 \um$) observations are ideal 
for quantifying the amount of AGN and star formation activity, respectively.
A key advantage of X-ray observations, specifically in the hard band (e.g.,
2 -- 8 keV), over other tracers of AGN activity, is that they 
are not greatly affected by the presence of obscuration and
contamination effects from the host galaxy (see sections 1 and 2 of 
\citealt{Alexander15} for more details of the use of the X-ray as an AGN tracer).  
A key advantage of FIR observations, as a measurement of star formation, 
is that they trace the peak of the 
obscured emission from star forming regions
 surrounded by cold gas and dust. Even though 
the FIR provides an indirect tracer of star formation,
a significant advantage over more direct tracers,
such as the UV and optical emission from the young massive stars,
is that it does not suffer significantly from obscuration
(e.g. \citealt{Kennicutt98}; 
\citealt{Calzetti07}; \citealt{Calzetti10}; see also section 2.2 in \citealt{Lutz14}).
Indeed, as shown by \cite{DominguezSanchez14}, for
luminous infrared galaxies (FIR luminosities of L$_{\rm FIR} \gtrsim$ 10$^{44} \ergss$)
more than 75\% of the total emission due to 
star formation is produced at FIR wavelengths, a fraction that increases 
at higher L$_{\rm FIR}$.\footnote{We note that for less luminous 
infrared galaxies (L$_{\rm FIR} \lesssim$ 10$^{44} \ergss$) 
\cite{DominguezSanchez14} find that the FIR emission accounts for $\sim$50\% 
of the total emission due to star formation. However in this work we find that our galaxies 
have average L$_{\rm FIR} \gtrsim$ 10$^{44} \ergss$ and so the majority of the star formation 
is expected to be produced at FIR wavelengths.} 
However, the AGN can also contribute
to the FIR luminosity due to the thermal re-radiation of 
obscuring dust from the surrounding torus (e.g. \citealt{Antonucci93}). 
Hence, for the most reliable measurements of the star formation it is
important to apply decomposition methods of the AGN
and star formation components at infrared wavelengths
(e.g. \citealt{Netzer07}; \citealt{Mullaney11}; \citealt{DelMoro13}; \citealt{Delvecchio14}).

%previous work

A number of studies have used X-ray and FIR observations to
understand the connection between distant AGN activity and star 
formation by measuring the mean SFRs of AGN and star forming 
galaxy samples
(e.g., \citealt{Lutz10};  \citealt{Shao10};
\citealt{Mainieri11}; \citealt{Mullaney12a};
 \citealt{Rovilos12}; \citealt{Santini12}; 
\citealt{Harrison12}; \citealt{Rosario13a, Rosario13b}; \citealt{Lanzuisi15}).
The main results shown by these studies are that: 
(1) the average star formation rates ($\langle$SFR$\rangle$) of AGN 
track the increase with redshift found for the overall star forming galaxy 
population; (2) the $\langle$SFR$\rangle$ of AGN are higher than those
of the overall galaxy population (i.e., when including quiescent galaxies);
and (3) the specific SFRs (i.e., the ratio of SFR over stellar mass, which 
serves as a measure of the relative growthrate of the galaxy) of AGN
are in quantitative agreement with those of star forming galaxies.
The majority of the current studies also
find no correlation between the AGN luminosity 
and $\langle$SFR$\rangle$ for moderate luminosity AGN 
(X-ray luminosities of $\lx \lesssim$ 10$^{44} \ergss$; 
e.g.,  \citealt{Lutz10};  \citealt{Shao10}; \citealt{Mullaney12a}; 
\citealt{Rovilos12}; \citealt{Harrison12}). 
However, there are significant disagreements in the results 
for high luminosity AGN ($\lx \gtrsim$ 10$^{44} \ergss$). 
There are studies arguing that the $\langle$SFR$\rangle$ 
increases at high AGN luminosities (e.g., \cite{Lutz10}; \citealt{Rovilos12}; \citealt{Santini12}),
a result that seems in agreement with the concept of AGN and star 
formation activity being connected due to their mutual dependence on 
the cold gas supply in the galaxy. Other studies have argued
that the SFR decreases at high AGN luminosities
(e.g., \citealt{Page12}; \citealt{Barger15}), potentially suggesting that the AGN may be responsible for reducing or even 
quenching the ongoing star formation (a result also inferred by some simulations
of galaxy evolution; e.g., \citealt{DiMatteo05}; \citealt{Hopkins05}; \citealt{Debuhr12}).
There are also studies arguing that  $\langle$SFR$\rangle$  remains constant  
up to high AGN luminosities (i.e., a broadly flat relationship; 
e.g., \citealt{Harrison12}; \citealt{Rosario12}; \citealt{Azadi14}),
extending the trend seen for moderate luminosity AGN.
Nevertheless, the difference in the conclusions of such studies 
could be attributed to the low source statistics for high luminosity 
AGN, and strong field to field variations (e.g., \citealt{Harrison12}).
For example, \cite{Harrison12} demonstrated
that when using a large high luminosity AGN sample the 
broadly flat relationship between $\langle$SFR$\rangle$  
and AGN luminosity found for moderate luminosity AGN 
continues to high luminosities, with no clear evidence for 
either a positive or negative correlation (see also \citealt{Harrison14} for a recent review).

%Outstanding questions/problems: why no positive correlation; AGN variability?
To first order a flat relationship between
 $\langle$SFR$\rangle$ and AGN
luminosity can seem surprising, 
since it appears to suggest the lack of a connection between 
AGN activity and star formation.
However, \cite{Hickox14} have shown that a
true underlying correlation between AGN luminosity 
and $\langle$SFR$\rangle$ can be masked if the
AGN varies significantly (i.e., by more than an order of magnitude)
on much shorter timescales 
than the star formation across the galaxy. 
In fact, observational studies such as 
\cite{Rafferty11}, \cite{Mullaney12b}, 
\cite{ChenTing13}, \cite{Delvecchio14}, and 
\cite{Rodighiero15} have shown that when the average 
AGN luminosity is calculated as a function of SFR
(i.e., taking the average of the more variable 
quantity as a function of the more stable quantity) 
a positive relationship is found,
suggesting that AGN activity and star formation are correlated on long timescales.
Studies using small scale hydrodynamical simulations of SMBH growth
(e.g., \citealt{Gabor13}; \citealt{Volonteri15}) have indeed suggested that AGN activity can vary
by a typical factor of $\sim 100$ over $\sim$Myr timescales, which results in 
a flat relationship between $\langle$SFR$\rangle$ and AGN
luminosity over a wide range of AGN luminosity. 
These studies therefore demonstrate that the relationship between AGN
luminosity and $\langle$SFR$\rangle$ can potentially place constraints on the
variability of mass accretion onto the SMBH in galaxies.
However, to date, the observational constraints of the $\langle$SFR$\rangle$ 
of AGN as a function of AGN luminosity and redshift
have lacked the accuracy to be able to distinguish between 
the different SMBH mass accretion models.

%Limitations of previous work
Most of the current studies on the $\langle$SFR$\rangle$  of distant X-ray AGN
suffer from a variety of limitations, which affect the accuracy 
of $\langle$SFR$\rangle$ measurements, such as: (1) small number of
sources, which can lead to large statistical uncertainties, 
particularly at high AGN luminosities; (2) high levels of 
source confusion at FIR wavelengths, which can 
cause the overestimation of the flux; (3) use of a single FIR
band from which to derive SFRs, which will result 
in large uncertainties on the $\langle$SFR$\rangle$ and will not take into 
account possible contamination of the SFR measurements from the AGN; 
(4) neglect of the information 
that can be obtained from the photometric upper
limits of the FIR undetected AGN, which make up the majority 
of the distant AGN in X-ray samples (this final point is not applicable for studies 
that use stacking analyses).

%Final paragraph: Summary and aim of paper
% why this work is better
In this work we aim to overcome the limitations outlined above by
exploiting a large sample of X-ray detected AGN with deep and extensive
multi-wavelength data, for which we perform
spectral energy distribution (SED) fitting on a source by source basis, 
and measure the SFR for each source
in our sample. We use deblended FIR photometry from \textit{Herschel}, 
which provides the best constraints on the FIR fluxes of individual sources 
by reducing the contamination due to 
blended and confused sources, the most significant drawback of the \textit{Herschel} 
field maps.
Furthermore, we make use of the photometric
upper limits in the fitting procedure to better constrain the SED
templates and SFRs. We finally calculate the $\langle$SFR$\rangle$ 
values as a function of
X-ray luminosity, with the inclusion of sources with only upper limit
constraints using survival analysis techniques (e.g., \citealt{Feigelson85}, 
\citealt{Schmitt85}). Our methods ensure the
use of all available data (i.e. photometric detections and upper
limits, SFR measurements and upper limits) 
to provide improved $\langle$SFR$\rangle$ 
as a function of X-ray luminosity and redshift. 
In Section \ref{data} we outline the photometric catalogues used in
this work, as well as the choice of redshift and the choice of matching
radii between photometric positions. In Section \ref{analysis} we
analyse our methods of SED-fitting as well as the calculation of the
average IR luminosity due to star formation ($\mlir$). Finally, in Section \ref{results}
we present and discuss our results. In our analysis we use $\rm H_0 = 71 km \, s^{-1}$,
$\rm \Omega_M = 0.27$, $\rm \Omega_\Lambda = 0.73$ and assume 
a \cite{Chabrier03} initial mass function (IMF).

\section{AGN Sample, IR photometry and Redshifts} \label{data}
In this work we use the available Mid--IR (MIR; $\lambda \approx 3 - 30 $)
to Far--IR (FIR;  $\lambda \approx 30 - 500 \um$) photometric
data to constrain the average SFRs of a large sample   
of  X-ray detected AGN over the redshift range $z \approx$ 0.2 -- 2.5. 
To construct a large sample of X-ray detected AGN we make use of 
three fields with deep X-ray observations:  
(1) \textit{Chandra} Deep Field North (CDF-N; \citealt{Alexander03}), (2)
\textit{Chandra} Deep Field South (CDF-S; \citealt{Xue11}), and (3) a
combination of \textit{Chandra}-COSMOS (C-COSMOS; \citealt{Elvis09}) and 
\textit{XMM}-COSMOS (\citealt{Cappelluti09}).
To construct our final AGN sample we obtain the MIR and FIR 
photometry from observations of the X-ray deep fields made with the 
\textit{Spitzer} (\citealt{Werner04}) and 
\textit{Herschel} (\citealt{Pilbratt10}) space observatories. 
The recent \textit{Herschel} observational programs PEP/GOODS-\textit{H}
(\citealt{Lutz11}; \citealt{Elbaz11}) and HerMES (\citealt{Oliver12})
in the three fields of GOODS-N, GOODS-S, and COSMOS,
covering the wavelength range of 70 -- 500$\um$ are our main source of
the FIR photometry (details in \S\ref{mfir}).
We therefore restrict the CDF-N and CDF-S X-ray catalogues
to these regions with sensitive MIR--FIR coverage, 
i.e. the GOODS-N and GOODS-S with areas of  
187\,arcmin$^2$ each, but use the full 2\,deg$^2$ of COSMOS.
In total these areas cover 3609 X-ray sources.
Figure \ref{lx_z} shows the X-ray sources in GOODS-N, GOODS-S,
C-COSMOS, and \textit{XMM}-COSMOS in the $\lx$ -- $z$ plane. 
	 
In the following subsections we describe our sample selection and
the catalogues used for the sample. In \S\ref{xray} we present
the X-ray observations used to define our AGN sample and to determine their
X-ray luminosities. In \S\ref{mfir} we present the
MIR and FIR photometric catalogues used to constrain the SFRs of
the AGN hosts via SED fitting. In 
\S\ref{matchcats} we describe the method of matching the X-ray sources
to the MIR and FIR catalogues and the redshift counterparts.

\subsection{X-ray Data} \label{xray}
\begin{figure}
  \begin{center}
    \includegraphics[scale=0.35]{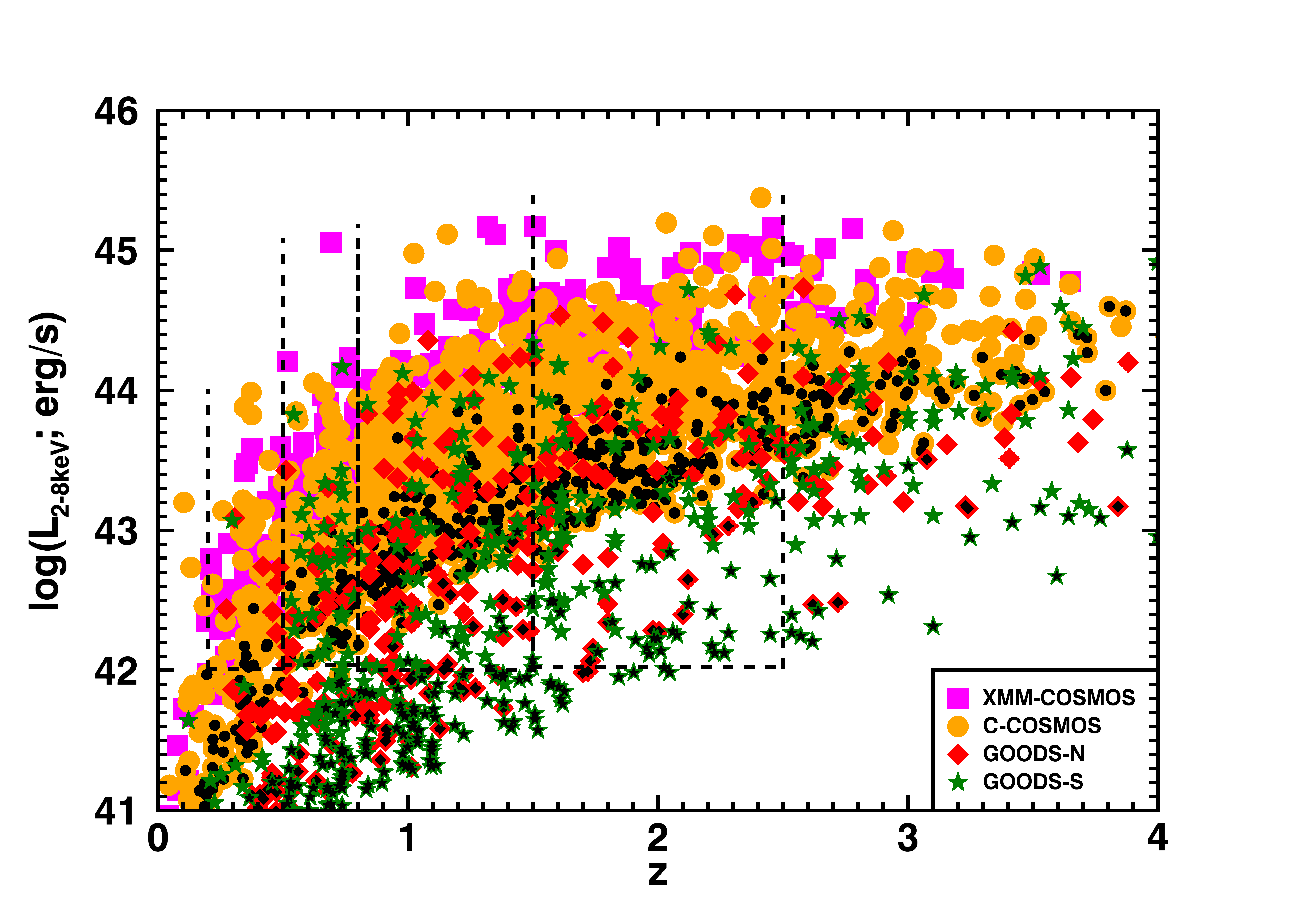} 
    \caption{X-ray (2-8 keV) luminosity ($\lx$) versus redshift ($z$) for the X-ray
      sources in the GOODS-N, GOODS-S, and COSMOS regions 
      described in \S\protect\ref{xray}. Black centers indicate the
      X-ray sources without a direct hard-band detection  
      (as described in \S\protect\ref{xray}). The vertical dashed lines indicate
      the 4 redshift ranges used in this study. The lower X-ray luminosity 
      threshold (L$_{\rm 2-8  keV} >$ 10$^{42}$ erg/s) used to define our AGN sample is 
      shown with the horizontal dashed line. The combination of the three fields 
      enables us to explore the SFRs of AGN over three orders of magnitude in AGN luminosity.} 
    \label{lx_z}
  \end{center}
\end{figure}

To select the sample of AGN for our study we use the publicly
available X-ray catalogues for the CDF-N (\citealt{Alexander03}),  
CDF-S (\citealt{Xue10}) and COSMOS (\citealt{Elvis09};
\citealt{Cappelluti09}) fields, restricted to the areas covered by
PEP/GOODS-\textit{H} and HerMES observations as described above 
(see Figure \ref{lx_z}). For the COSMOS
field we use the C-COSMOS X-ray catalogue as the primary sample,
while for the sources over the larger region, not covered by \textit{Chandra}, we use
the \textit{XMM}-COSMOS catalogue. 
Rest-frame, hard band 2 -- 8\,keV luminosities were
calculated following \cite{Alexander03a} with the equation: \begin{equation} 
\lx= \rm 4\pi \times D_L^2 \times F_{2-8\rm{keV}} \times (1+z)^{(\Gamma-2)}
\end{equation}  
where F$_{\rm 2-8keV}$ is the observed X-ray hard band flux (2--8 keV), 
D$_{\rm L}$ is the luminosity distance, $z$ is the redshift (see Section
\ref{matchcats}), and $\Gamma$ is the photon index used for
k-corrections, which was fixed to a standard value of $\Gamma=1.9$
(e.g., \citealt{Nandra94}). Although the hard band observations in CDF-N and CDF-S are in
the energy range of 2--8 keV, the C-COSMOS and \textit{XMM}-COSMOS catalogues
report hard band fluxes of the energy range of 2--10 keV. To
convert the 2--10 keV to 2--8 keV fluxes we assume $\Gamma=1.9$ which
yields a conversion factor of 0.85. 

For the 20\% of X-ray sources in our final sample (see below) not detected in the hard band we used the full
band of 0.5 -- 8\,keV (or the soft band of 0.5 -- 2\,keV if undetected 
in the full band) to estimate the hard band flux. 
We estimated the hard-band flux assuming a $\Gamma =$ 1.4 spectral slope, unless 
this provided a measurement greater than the hard-band upper limit, in which case 
we assumed a $\Gamma =$2.3 spectral slope; the assumed range in spectral 
slope is motivated by the range observed in AGN (e.g., \citealt{Nandra94}; 
\citealt{George00}). Overall, with this procedure, the hard-band fluxes were estimated 
assuming $\Gamma =$ 1.4 for 19\% and $\Gamma =$ 2.3 for 1\% of sources 
in our sample (see Figure \ref{lx_z}).

\subsection{Mid-IR \& Far-IR Data} \label{mfir}
To measure the SFRs of our AGN sample we need to
reliably constrain the IR luminosity due to star formation and
remove any contribution from the AGN. To do this we need data covering both  
the MIR and FIR wavelengths for
each source in our sample (e.g., \citealt{Mullaney11}). We
exploit available photometry in the wavelength range of 8$\um$ -- 500$\um$,
provided by observations carried out by: \textit{Spitzer}-IRAC at
8$\um$; \textit{Spitzer}-IRS at 16$\um$; \textit{Spitzer}-MIPS at
24$\um$, 70$\um$; \textit{Herschel}-PACS at 70, 100, 160$\um$; and 
\textit{Herschel}-SPIRE at 250, 350, 500$\um$.
One of the advantages of our study over several previous studies,
is the use of catalogues of deblended FIR \textit{Herschel} images (details below). 
The deblending of sources in the PACS and SPIRE 
observations allows us to overcome the blending and confusion issues
encountered in dense fields that can lead to an overestimation of  
the flux densities (e.g., \citealt{Oliver12}; \citealt{Magnelli13}). 
It also ensures the direct association between the measured FIR
flux densities and the sources used as priors in the deblending
process. In addition to this, we also make sure that we have a reliable
photometric upper limit for sources not
detected in the FIR. This enables us to constrain the star forming galaxy
templates and gain an upper limit on the IR luminosity due to star 
formation, as we describe in \S\ref{sed_an}. 
 
The MIPS 24$\um$ photometric catalogues that we use were created by
\cite{Magnelli13}. 
These catalogues are made by simultaneous PSF 
fitting to the prior positions of 3.6$\um$ sources. 
The catalogues were limited to a $3\sigma$ detection limit at 24$\um$ 
going down to 20$\rm \mu Jy$ in GOODS-N and
GOODS-S, and 50$\rm \mu Jy$ in COSMOS.  
The PACS 70$\um$, 100$\um$ and 160$\um$ catalogues were also created by
\cite{Magnelli13} using the MIPS 24$\um$ detected sources,  
described above, as the priors for the deblending of the PACS
maps. Only sources with at least a $3\sigma$ detection at MIPS 24$\um$
were used as priors and the resulting PACS catalogues were also limited to a
$3\sigma$ detection limit. \footnote{The PACS catalogues for GOODS-N and GOODS-S are 
published in \cite{Magnelli13}. The catalogue for COSMOS was created in the 
same way and is available online (http://www.mpe.mpg.de/ir/Research/PEP/DR1).}  
The SPIRE 250$\um$, 350$\um$, and 500$\um$  
catalogues were created following the method described in
\cite{Swinbank14}, again using these MIPS 24$\um$ positions as priors to
deblend the SPIRE maps.  
 
 Although both the PACS and SPIRE catalogues have been produced in 
the same way, \cite{Magnelli13} do not provide flux upper limits. 
In order to keep the priored FIR catalogues
consistent with each other, we calculate upper limits for the  
non-detections in the PACS catalogues of \cite{Magnelli13} in a
similar way to the upper limit calculation performed for the SPIRE  
priored catalogues of \cite{Swinbank14}. This was done by performing
aperture photometry at thousands of random positions in the PACS
residual maps and taking the 99.7th
percentile of the distribution of the measured flux densities as the
$3\sigma$ upper limit on the nondetections. To account for the fact
that the deblending is more uncertain in regions of luminous sources,
we calculated these $3\sigma$ upper limits as a function of the pixel
values in the original maps (see \citealt{Swinbank14}). Consequently,
this approach results in upper limits being higher for non-detected
sources that lie near a bright source, when compared to non-detected sources in
blank areas of the maps.  

Due to the fact that we are using MIPS 24$\um$ priored 
catalogues for the FIR photometry of our sources, any  
undetected at 24$\um$ will not have FIR measurements in the published catalogues.
Therefore for the  24$\um$ undetected sources, we extracted the FIR photometry at the optical counterpart positions 
following the method described in \cite{Swinbank14}. 
Overall there are only 23 sources that are undetected at  24$\um$ but have 
FIR counterparts, making up a very small fraction of our overall sample.

In the MIR bands we also use the catalogues of \textit{Spitzer}--IRAC 8$\um$ observations 
as described in \cite{Wang10}, \cite{Damen11}, and \cite{Sanders07}, for
GOODS-N, GOODS-S, and COSMOS, respectively, as well as \textit{Spitzer}--IRS 16$\um$
from \cite{Teplitz11} for GOODS-N and GOODS-S. Since all
the IRAC catalogues have their detections determined by the 3.6$\um$
maps, and the 16$\um$ catalogues have been produced with the use  
of 3.6$\um$ priors, they are all consistent with the deblended PACS and SPIRE 
catalogues described above.

% Table with the total number of AGN in each field, number of sources with at least 24um counterpart, number of sources with redshift+spec. redshift. 
\begin{table}
  	\begin{center}
    	\begin{tabular}{|lcccc|}  \hline \hline
      	Field & AGN & with spec-$z$ &  with 24$\um$ \\
      	\hline
      	GOODS-N   & 177 & 98 & 137\\
      	GOODS-S   &  209 & 128 & 154\\	
      	COSMOS    & 1753 & 914 & 1151\\
      	\hline
      	Total     & 2139 & 1140 & 1442\\
    \end{tabular}
    \caption{Number of X-ray detected AGN in our parent sample
      (L$_{\rm 2-8 keV}$ $>$ 10$^{42}$ erg/s; $z =$ 0.2--2.5)  
      in each field, as well as the number of sources with a
      spectroscopic redshift and the number of sources with a $24\um$
      counterpart.} 
    \label{tabinfo}
  \end{center}
\end{table}

\subsection{Redshifts and catalogue matching}	 \label{matchcats}

For our SED fitting analysis (see \S\ref{sed_an}) we need matched
catalogues containing X-ray fluxes, MIR-FIR photometric flux densities, and
redshifts. To obtain the appropriate counterparts for each X-ray
source, we matched the catalogues starting with the X-ray catalogues described in 
\S\ref{xray}. We first match the positions of the optical counterparts of the X-ray sources
to the MIPS 24$\um$ positions in the catalogues of \cite{Magnelli13}.\footnote{
For the X-ray catalogues of CDF-S and C-COSMOS the optical
counterparts are provided by \cite{Xue10} and \cite{Elvis09}. For the sources  
in CDF-N we use the catalogue of \cite{Barger08}.}
To choose the matching radii between catalogues we measure the number
of total matches as a function of radius and estimate the fraction of
spurious matches for each matching radius. The matching 
radius of the X-ray to the MIPS 24$\um$ catalogue for GOODS-N and GOODS-S 
was 0.8", while for C-COSMOS and \textit{XMM}-COSMOS it was 1".
This matching radius was chosen to
maximise the number of matches while minimising the number of spurious
matches, with a ratio of spurious to true matches of 1\%. 
Due to the way that the FIR catalogues were deblended, each MIPS 24$\um$
detected source also has a corresponding photometric measurement or flux 
upper limit for PACS 70$\um$, 100$\um$, 160$\um$ and SPIRE
250$\um$, 350$\um$, 500$\um$ (see \S\ref{mfir}).  For the sources not
matched to a MIPS 24$\um$ counterpart we use the FIR data extracted 
at their optical counterpart positions, as described in \S\ref{mfir}.
We then match to the IRAC, and to the IRS 16$\um$ catalogues for
the two GOODS fields (see \S\ref{mfir}) using the same method. 

A necessity for this analysis are the redshifts of the X-ray sources.
To allocate the redshift counterpart of the sources in GOODS-S 
and C-COSMOS we make use of the 
spectroscopic and photometric redshift
compilation by \cite{Xue11} and \cite{Civano12},  
respectively. We also added redshifts from \cite{Teplitz11} for
sources in GOODS-S when necessary. For the sources in GOODS-N we created  
our own compilation using catalogues of spectroscopic redshifts from
\cite{Barger08} and \cite{Teplitz11} and photometric  
redshifts from \cite{Wirth04} and \cite{Pannella09}. Overall we
obtained redshifts for 91.4\% of the X-ray sources. 

In total there are 3297 X-ray sources covered by \textit{Chandra}, \textit{XMM}, and
PEP/GOODS-\textit{H} observations with a   
redshift (see Figure \ref{lx_z}). For this study we restrict this sample to redshifts of $z
=$ 0.2 -- 2.5 and a luminosity range of $\lx$ $>$ 10$^{42} \ergss$, 
resulting in our parent sample of 2139 AGN.  
Of the parent sample 53.3\% have spectroscopic redshifts and 67.4\%
are detected at MIPS-24$\um$ (see Table \ref{tabinfo} for a
summary of the three fields).

%%%%%%%%%%%%%%%%%%%%%%%%%%%%%%%%%%%%%%%%%%%%%%%%%
\section{Data analysis} \label{analysis}
In this study we are interested in measuring the mean SFRs of galaxies,
hosting an X-ray detected AGN, as a function of the AGN luminosity and redshift.   
We use multi-band IR photometry, including photometric upper limits, to perform SED fitting  
for all 2139 X-ray detected AGN in our parent sample (see 
\S\ref{matchcats}; Figure \ref{lx_z}). For each source we decompose the
contribution of AGN activity and star formation to the overall SED.
This allows us to measure the IR luminosity due to star formation ($\lir$),
the key quantity for this study, which we can use as a proxy for SFR 
(e.g., \citealt{Kennicutt98}, \citealt{Calzetti07}, \citealt{Calzetti10}).
In \S\ref{sed_an} we outline the SED fitting procedure and
describe the calculation of $\lir$. In \S\ref{meantech}
we describe the method that we follow for the calculation of
the average $\lir$ as a function of $\lx$ (our tracer of the AGN luminosity) for the whole
sample, where we include both direct $\lir$ measurements and upper
limits. The calculation of
these values thus allows us to investigate how SFR relates to AGN
luminosity (Section \ref{results}).

\subsection{SED fitting procedure} \label{sed_an}
\begin{figure*}
		\begin{center}
			    \includegraphics[scale=0.6]{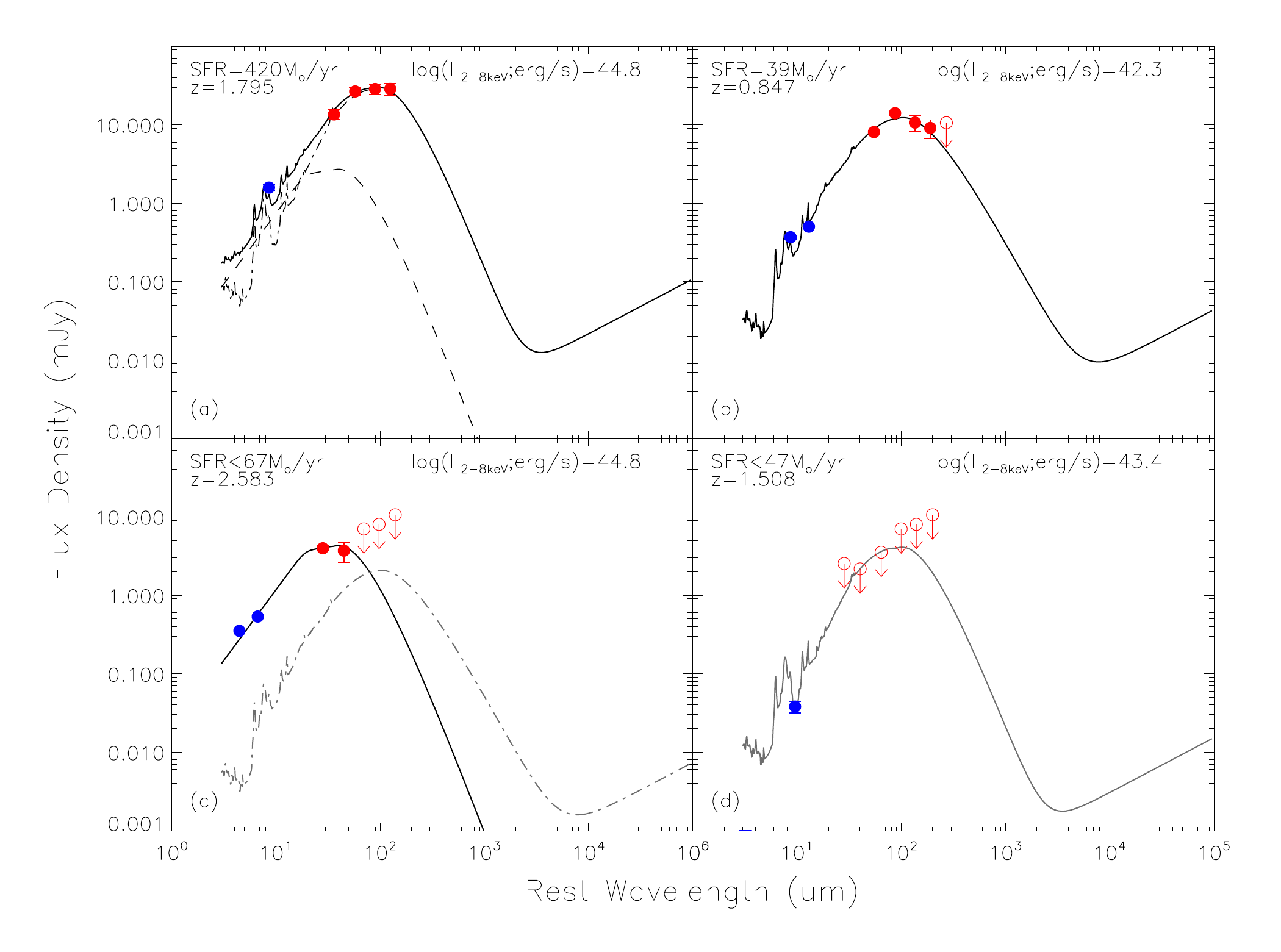}		
				\caption{Examples of the four types of best fitting SED solutions. 
				(a) A galaxy where the best fit (solid curve) is the combination 
                 of AGN (dashed) and star forming galaxy (dot-dashed curve) templates. (b) A galaxy where 
                 the best fit is that of a star forming galaxy template alone, with 
                 no AGN contribution. (c) A galaxy where the best fit solution is an 
                 AGN (solid curve) with no star formation contribution, in this case we calculated 
                 an upper limit on the star forming component shown by the grey dot-dashed curve. 
                 (d) A galaxy with only one photometric detection where we can only calculate 
                 an upper limit for the star forming galaxy templates, as shown here by the grey curve.
                 In all four cases the blue data points are from \textit{Spitzer} observations, 
                 while red data points are from \textit{Herschel} observations. The filled 
                 circles are measured flux densities, while the empty circles with an arrow are the flux
                 density upper limits. For each example we also give the SFR, X-ray luminosity, and 
                 redshift of the source.
                 The wavelengths have been shifted to the rest frame, but the flux densities
                 are in the observed frame.}
				\label{sed_examples}
		\end{center}	
	\end{figure*}

To calculate individual $\lir$ values for our sample we perform SED fitting to
the MIR and FIR photometry. In these bands there could be a
contribution from both AGN and star formation, with emission from the AGN peaking
at MIR wavelengths and dropping off at the FIR wavelengths 
(e.g., \citealt{Netzer07}; \citealt{Mullaney11}). Those factors make it
important to decompose the contribution from both star formation and AGN to the
overall SED so as to avoid an overestimation of the SFR measurement.
In Figure \ref{sed_examples} we give example SED
fits to demonstrate our procedure.

To fit and decompose the IR SED of our sources we develop the publicly
available DecompIR code of \cite{Mullaney11}, and use the 
8 -- 500$\um$ data and upper limits described in \S\ref{mfir}. 
We use a set of empirical templates that consist of the mean AGN template 
and the five star forming galaxy templates originally defined in \cite{Mullaney11}, and
extended by \cite{DelMoro13} to cover the  
wide wavelength range of 3 -- 10$^5$ $\um$. We also include the Arp220
galaxy template from \cite{Silva98} which serves  
as a sixth template to ensure that we are also covering the
possibility of extremely dusty star forming systems. The advantage of using  
a few, but representative, templates to fit the data is that
we can avoid the degeneracy in the fitting procedure caused  
by a large number of templates. Furthermore as many of our sources have
limited photometric detections (with only one or two data points), 
it is sensible to keep the number of free parameters as small as
possible. We note that the set of star forming galaxy templates described above  
covers a broad range of empirical shapes, including the large template
library of \cite{Chary01}, as shown in Figure 2  
of \cite{DelMoro13}, and the templates described by
\cite{Kirkpatrick12}.

In our fitting procedure the only free parameters of the fit are the
normalisation of the star forming galaxy and AGN templates. Since there are two free 
parameters in the fit we require that the source has at least three photometric
detections to simultaneously fit the AGN and star forming galaxy templates.  
When we have less than three photometric detections we can only derive
upper limits on $\lir$, as we cannot constrain the AGN contribution (see below).

When a source is detected in three or more photometric bands we perform a
series of fits following the method of \cite{DelMoro13}.  
We fit the data in two steps: firstly we fit using each of the six star forming
galaxy templates separately without including the AGN component,
and secondly we fit again with each of the star forming galaxy templates in combination 
with the AGN template. We fit to the photometric flux density detections, 
but use the available flux density upper limits to eliminate the fits
which are above any of the upper limits. This procedure
results in a maximum of twelve models (the six star forming 
galaxy templates without an AGN and
the six star forming galaxy templates with an AGN) to chose from. 
 
To determine the best fitting solution of the twelve possibilities described above, we
use the Bayesian Information Criteria (BIC; \citealt{Schwarz78}) 
which allows the objective comparison of different non-nested models
with a fixed data set, and is defined as: \begin{equation}
\rm BIC = -2\times lnL + k\times lnN
\end{equation}
where L is the maximum likelihood, k is the number of free parameters,
and N the number of data points. This method penalises against models with extra
free parameters counterbalancing the fact that a model with more free parameters 
can fit the data better, irrespective of the relevance of the parameters. This 
is an improvement over a simple $\rm \Delta\chi^{2}$ test or a maximum likelihood 
comparison that would tend to favour the model with more free parameters. 
For each source the BIC value
is calculated for all of the different fits. 
The best fitting model will be the one
which minimises the BIC value, its absolute value being irrelevant;
however for one model to be significantly better than the others it needs
to have a difference in BIC value of $\rm \Delta BIC \geq$ 2. If $\rm \Delta BIC \leq$ 2
then both models are considered equally valid (e.g. \citealt{Liddle04}).  
Our final best fit solution is the one
with the lowest BIC value; however we only accept the AGN component as 
significant if the inclusion of it reduces the BIC value by $\geq$ 2.
In Figure \ref{sed_examples}(a) we show a best fit SED that includes 
the AGN and star formation component, and in Figure \ref{sed_examples}(b) 
a best fit SED with only the star formation component.
From the best fit SEDs we then measure the integrated $8 - 1000\um$ 
IR luminosity of the star formation component ($\lir$).
Furthermore, if multiple fits have BIC values equal to the minimum
BIC value, we consider them equally valid and take the average of their 
derived $\lir$.

For sources detected in fewer than three photometric bands we can only
calculate upper limits on $\lir$, due to the  
insufficient degrees of freedom to calculate the AGN contribution to the 
IR luminosity. To calculate the upper limits of the
normalisation of each star forming galaxy template we increase the normalisation of each template 
until it reaches one of the 3$\sigma$ upper limits, or exceeds the
3$\sigma$ uncertainty of a data point. We take the star forming galaxy template with the highest
upper limit of $\lir$ as our conservative upper limit for that source
(e.g. see Figure \ref{sed_examples}(d)). Using the same method
we also derive upper limits on the star formation contribution for sources where
the best fit is fully dominated by the AGN (e.g. see Figure \ref{sed_examples}(c)). 

Due to the limited photometry and quality of the data, our procedure is not expected to significantly detect an AGN 
component in the IR SEDs of all sources. Indeed, the detection of the AGN component in the MIR will be dependent 
on the relative ratio of $\lir$ over the IR luminosity due to the AGN ($\rm L_{IR,AGN}$); for example,
a source with a high ratio of $\lir$ over $\rm L_{IR,AGN}$ will not show strong evidence of an AGN 
component in its IR SED (e.g., see Appendix A of \citealt{DelMoro13}).
However, we note that if we force an AGN component to be present in the IR SEDs of each of our 
sources, our results of mean $\lir$ in bins of X-ray luminosity and redshift (see \S \ref{meantech})
only change within a $\sim$5\% level, which is smaller than the uncertainty of the 
mean $\lir$ results presented in \S \ref{meansfr_lx}.
We also verified that our results were not sensitive to the choice of AGN template that we used.
by refitting sources with two different AGN templates. 
One template is representative of low luminosity AGN, while the other template is representative 
of high luminosity AGN, as provided by \cite{Mullaney11}.
The first template is ``colder" than that used in our main analysis, 
with less emission in the MIR and extended emission to the FIR wavelengths, and the second template is ``hotter", 
with most emission occurring at MIR wavelengths and a steep drop-off in the FIR
(in agreement with the mean empirical templates of Quasars in the FIR; e.g. \citealt{Netzer07}).
Between them, these two templates, encompass most clumpy-torus models (see Fig. 7 in \citealt{Mullaney11}). 
In both cases our results of mean $\lir$ in bins of X-ray luminosity and redshift (see \S \ref{meantech})
only change within a $\sim$10\% level, which again is smaller than the uncertainty in the 
mean $\lir$ results presented in \S \ref{meansfr_lx}.

Using our SED fitting approach we have a sample of 2139 AGN with individual
measurements (including upper limits) of $\lir$. From our results for the whole sample
there are 263 fits that required a significant AGN component in addition to star formation, 274 fits that required only the 
star forming galaxy template, and for 1602 sources only upper limits on the star formation component 
could be derived due to limited photometry. 

\subsection{Calculating average source properties} \label{meantech}
For this study we aim to constrain the average star formation rates
of our X-ray AGN sample as a function of redshift and X-ray luminosity.  
A challenge for all studies using \textit{Herschel} 
FIR photometry is the low detection rate of individual sources (e.g., \citealt{Mullaney12a}). 
In our sample we can only place upper limit
constraints on the $\lir$ for many of our sources, i.e. 1612 out of
the 2139 (75.4\%) sources in our sample, due to the limited photometry 
or because they are AGN dominated.
In our study we have placed conservative upper 
limits on the $\lir$ for the AGN for which it was not 
possible to directly identify the star formation component (see \S\ref{sed_an}). 
In order to not bias our study to only the FIR
bright sources we study the average properties of the
whole X-ray selected AGN sample by using a Survival Analysis
technique (e.g., \citealt{Feigelson85}, \citealt{Schmitt85}) to calculate the mean IR
luminosities with the inclusion of all of the upper limits (details below).  

We divide our sample in to four
redshift ranges, $z =$ 0.2 -- 0.5, 0.5 -- 0.8, 0.8 -- 1.5, and 1.5 -- 2.5. 
For each redshift range we also divide the sample 
in to L$_{\rm{2-8keV}}$ bins determined such that they included 
$\approx$40 sources in each bin. To ensure that all of the
 sources within the redshift range are 
included we allow the number to vary slightly, resulting in bins of 40 -- 43 sources.
For each $\lx$ -- $z$ bin we 
calculate the mean IR luminosity due to star formation ($\mlir$; see \S\ref{sed_an}) 
and mean X-ray luminosity ($\mlx$; see \S\ref{xray}). To calculate the 
$\mlir$ values, with the inclusion of upper limits, we use the
Kaplan-Meier product limit estimator (\citealt{KaplanMeier58}), a non-parametric 
maximum-likelihood-type estimator of the distribution function. We
use the formula as described in \cite{Feigelson85} for 
the estimation of the mean of a sample including the upper limit
values. The advantage of this method is that it does 
not assume an underlying distribution. We will refer to this method as the  
K-M method for the rest of this paper. 
 
The main requirement for the use of the K-M method,
is for the upper limit values to be randomly 
distributed among the measured values and independent of them.
Due to the different types of upper limits that result 
from our fitting procedure (see \S\ref{sed_an}) the upper limits on $\lir$ are indeed random enough for the use of this 
method. \footnote{Our SED fitting procedure provides upper limits for the cases where a source is:
(a) MIR -- FIR undetected; (b) MIR -- FIR detected, but in less than three bands; (c) AGN dominated, 
i.e. the fit doesn't require any contribution from the SF templates. In the case of (a) the upper limits are calculated 
by constraining the SF templates to the flux upper limits, while in the cases of (b) and (c) they are calculated by 
constraining the SF templates to the 3$\sigma$ flux errors or the flux upper limits. The fact that bright IR sources can 
meet the criteria of (b) and (c), in combination with the spatial dependence of the FIR flux upper limits, helps
drive the similarity between the distributions of the $\lir$ upper limits and measurements.}
Furthermore, a K-S test on our $\lx$ -- $z$ bins, with a probability threshold of 1\%, 
shows no evidence of the distributions of upper limits
and measured values being drawn from different distributions.
This method also requires that the lowest $\lir$ value in each bin is a measurement and not 
an upper limit. For the 12 bins where this is not the case we follow the popular procedure 
amongst studies using this method, and assume that the lowest value is a measurement
(e.g., \citealt{Feigelson85}, \citealt{Zhong09}). 
These 12 bins are randomly distributed with $\lx$ and redshift (see Table \ref{tabresults}) , and therefore do not 
affect our conclusions on the trends of $\mlir$ with redshift, and $\lx$.

 \cite{Feigelson85} use the K-M method to estimate means with up to a censorship 
(i.e., the fraction of upper limits) of 90\%, but argue that there can be a significant 
bias in such cases. Additionally, a study by \cite{Zhong09} estimating the bias of this 
method for a wide range of distribution types, find that
the estimated means are within a factor of 2 for up to 80--90\% censorship levels.
In our work we have imposed a limit of 90\%  censorship on our bins, and have discarded 7 bins 
with greater censorship. The median censorship level amongst 
the remaining 45 bins we have used in our analysis is $\sim$73\%, 
with 11 of them having a censorship of 80--90\% (see Table \ref{tabresults}).
For the calculation of the uncertainty on the mean we use the 
bootstrap technique, for which we take 10000
random samplings in each bin and recalculate the mean.  
We then take the 16th and 84th percentiles of the overall distribution as
the 1$\sigma$ errors. As discussed above, bins of high censorship levels could
suffer from additional uncertainties of a factor of $\lesssim$2. However, when comparing 
to the results of the stacking procedure, we find that the two methods are consistent (see Appendix), 
and hence, we do not have concerns about the high censorship levels in our bins causing a significant systematic bias.

We show our final results of $\mlir$ as a function of $\lx$ in Figure \ref{meanLirLagn}.
In our plots, throughout Section \ref{results}, we also include axes of SFR and AGN bolometric 
luminosity ($\lagn$) to help interpret the $\lir$ and $\lx$ measurements. We calculate $\lagn$ from $\lx$ by using the luminosity dependent
relation of \cite{Stern15} to convert the $\lx$ to an AGN 6$\um$
luminosity density. We then multiply this by a factor of 8 to
convert the 6$\um$ luminosity density to $\lagn$ (following \citealt{Richards06}). 
The SFRs were calculated from the $\mlir$ with the use of the \cite{Kennicutt98} 
relation corrected to a Chabrier IMF (\citealt{Chabrier03}).
	
%%%%%%%%%%%%%%%%%%%%%%%%%%%%%%%%%%%%%%%%%%%%%%%%%	
\section{Results \& Discussion} \label{results}
In this section we present our results and explore the form of the 
relationship between the average SFR, $\mlir$, and X-ray luminosity, $\lx$, 
for our sample of 2139 X-ray detected AGN (see Section \ref{data}) .
In \S\ref{meansfr_lx} we present our results of average SFR
(calculated from $\mlir$) as a
function  of X-ray (and bolometric) AGN luminosity for four redshift
ranges within $z =$ 0.2 -- 2.5. In \S\ref{MS} we compare the 
SFR of the AGN to those of the overall star forming galaxy population, for a
subsample of our sources with reliable host-galaxy masses. In \S\ref{models} we compare our results to
the predictions from two empirical models that connect AGN activity to
star formation. 

\subsection{Mean star formation rate as a function of X-ray luminosity} \label{meansfr_lx}
\begin{figure*} 
			\begin{center}
				\includegraphics[scale=0.5]{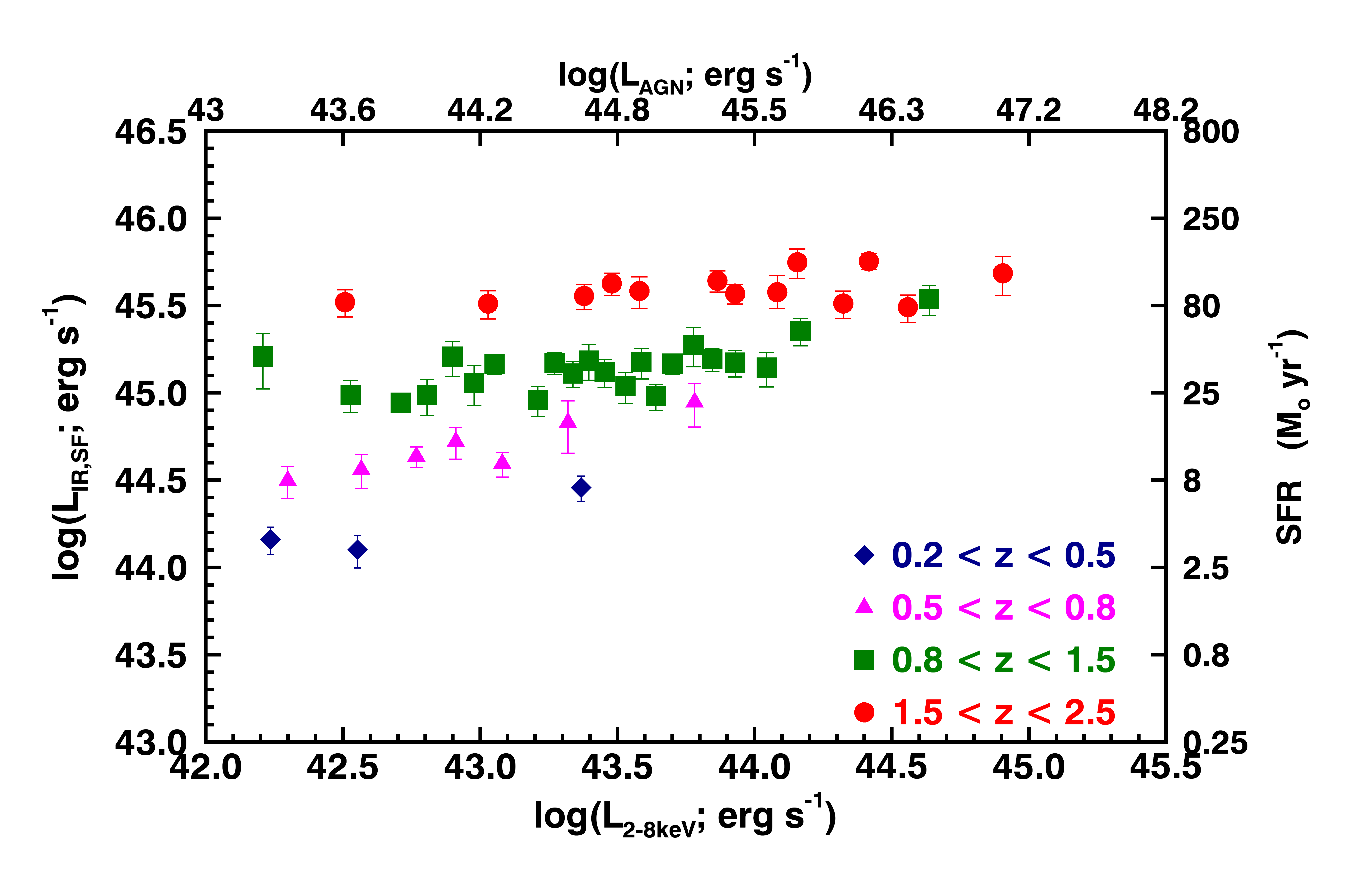}
 				\caption{Mean IR luminosity due to star formation,
                                  $\mlir$, as a function of X-ray
                                  luminosity, $\mlx$, for four
                                  redshift ranges. Each
                                  $\lx$ bin has $\sim$40
                                  sources. We also give the
                                  corresponding SFR values using the
                                  \protect\cite{Kennicutt98} relation
                                  corrected to a Chabrier IMF
                                  (\protect\citealt{Chabrier03}), and the
                                  bolometric AGN luminosity $\lagn$ calculated from 
                                  $\lx$ using the luminosity dependent relation 
                                  of \protect\cite{Stern15}. The errors  
                                  on the $\mlir$ are calculated using the
                                  bootstrap analysis as described in
                                  \S\protect\ref{meantech} (see also \S
                                  \protect\ref{meantech} for a discussion on
                                  the additional uncertainties).}
				\label{meanLirLagn}
			\end{center}
	\end{figure*}

The main focus of this paper is to determine the form of 
the relationship between the average SFR 
and AGN X-ray luminosity over 4 redshift ranges. 
The results of our analysis as described in \S\ref{meantech}
are presented in Figure \ref{meanLirLagn} and Table
\ref{tabresults}. In Figure \ref{meanLirLagn} the data are colour coded
by redshift where each point is the mean of $\approx$40 sources
and the error bars correspond to the 1$\sigma$ of the bootstrap 
errors (see \S\ref{meantech}).
We find that the $\mlir$ (and hence $\langle$SFR$\rangle$)
increases with redshift, by a factor of $\sim$3
between each redshift range, in agreement with both the observed evolution
found for normal star forming galaxies (e.g. \citealt{Elbaz11}; \citealt{Schreiber14})
and previous studies on AGN populations
(e.g.  \citealt{Shao10}; \citealt{Rovilos12}; \citealt{Rosario12}; \citealt{Mullaney12a}). 
However for the individual redshift ranges we find no strong correlation between 
$\mlir$ and $\lx$, a result inconsistent 
with that suggested by some other studies which have reported a rise or fall of 
$\mlir$ at high X-ray luminosities (e.g., \citealt{Lutz10}; \citealt{Page12}; \citealt{Rovilos12}; although see \citealt{Harrison12}).
 
\begin{table*}
  	\begin{center}
\begin{tabular}{|c|c|c|c|c|c|c|c|}
\hline
  \multicolumn{1}{|c|}{$\langle$z$\rangle$} &
  \multicolumn{1}{c|}{$\mlx$} &
  \multicolumn{1}{c|}{$\mlagn$} &
   \multicolumn{1}{c|}{Censorship} &
  \multicolumn{1}{c|}{$\mlir$} &
  \multicolumn{1}{|c|}{$\langle$SFR$\rangle$} &
 \multicolumn{1}{c|}{flag} \\
\hline
	& $\ergss$ & $\ergss$ & \% & $\ergss$ & $\rm M_{\odot}/yr^{-1}$  & \\
  \hline
  0.38 & 1.7$\times$10$^{42}$ & 1.9$\times$10$^{43}$ & 70 & 1.4$^{+0.3}_{-0.3}\times$10$^{44}$ & 4$^{+1}_{-1}$ & 1\\
  0.36 & 3.6$\times$10$^{42}$ & 4.5$\times$10$^{43}$ & 72 & 1.3$^{+0.3}_{-0.3}\times$10$^{44}$ & 3$^{+1}_{-1}$ & 0\\
  0.39 & 2.3$\times$10$^{43}$ & 6.9$\times$10$^{44}$ & 54 & 2.9$^{+0.5}_{-0.5}\times$10$^{44}$ & 8$^{+1}_{-1}$ & 0\\
  \hline 
  0.65 & 2.0$\times$10$^{42}$ & 2.3$\times$10$^{43}$ & 70 & 3.1$^{+0.6}_{-0.7}\times$10$^{44}$ & 8$^{+2}_{-2}$ & 0\\
  0.68 & 3.7$\times$10$^{42}$ & 4.6$\times$10$^{43}$ & 79 & 3.6$^{+0.8}_{-0.8}\times$10$^{44}$ & 10$^{+2}_{-2}$ & 1\\
  0.66 & 5.9$\times$10$^{42}$ & 7.9$\times$10$^{43}$ & 74 & 4.3$^{+0.6}_{-0.6}\times$10$^{44}$ & 11$^{+2}_{-2}$ & 1\\
  0.68 & 8.2$\times$10$^{42}$ & 1.2$\times$10$^{44}$ & 67 & 5.2$^{+1.1}_{-1.1}\times$10$^{44}$ & 14$^{+3}_{-3}$ & 0\\
  0.67 & 1.2$\times$10$^{43}$ & 1.9$\times$10$^{44}$ & 70 & 3.9$^{+0.6}_{-0.6}\times$10$^{44}$ & 10$^{+2}_{-2}$ & 0\\
  0.68 & 2.1$\times$10$^{43}$ & 3.9$\times$10$^{44}$ & 67 & 6.8$^{+2.2}_{-2.2}\times$10$^{44}$  & 18$^{+6}_{-6}$ & 0\\
  0.67 & 6.0$\times$10$^{43}$ & 1.8$\times$10$^{45}$ & 61 & 8.8$^{+2.4}_{-2.4}\times$10$^{44}$ & 23$^{+6}_{-7}$ & 0\\
  \hline 
  1.11 & 1.6$\times$10$^{42}$ & 1.8$\times$10$^{43}$ & 48 & 1.6$^{+0.6}_{-0.6}\times$10$^{45}$ & 43$^{+15}_{-15}$ & 0\\
  1.04 & 3.4$\times$10$^{42}$ & 4.2$\times$10$^{43}$ & 75 & 9.7$^{+2.0}_{-2.0}\times$10$^{44}$  & 26$^{+5}_{-5}$ & 0\\
  1.02 & 5.1$\times$10$^{42}$ & 6.8$\times$10$^{43}$ & 70 & 8.8$^{+1.0}_{-1.0}\times$10$^{44}$ & 23$^{+3}_{-3}$ & 0\\
  1.0 & 6.4$\times$10$^{42}$ & 8.8$\times$10$^{43}$ & 68 & 9.7$^{+2.2}_{-2.3}\times$10$^{44}$ & 26$^{+6}_{-6}$ & 0\\
  1.1 & 7.9$\times$10$^{42}$ & 1.1$\times$10$^{44}$ & 65 & 1.6$^{+0.4}_{-0.4}\times$10$^{45}$ & 43$^{+10}_{-10}$ & 0\\
  1.1 & 9.5$\times$10$^{42}$ & 1.4$\times$10$^{44}$ & 73 & 1.1$^{+0.3}_{-0.3}\times$10$^{45}$ & 30$^{+8}_{-8}$ & 0\\
  1.09 & 1.1$\times$10$^{43}$ & 1.8$\times$10$^{44}$ & 73 & 1.5$^{+0.2}_{-0.2}\times$10$^{45}$  & 39$^{+5}_{-5}$ & 1\\
  1.07 & 1.6$\times$10$^{43}$ & 2.8$\times$10$^{44}$ & 88 & 9.0$^{+1.8}_{-1.8}\times$10$^{44}$ & 24$^{+5}_{-5}$ & 0\\
  1.15 & 1.9$\times$10$^{43}$ & 3.3$\times$10$^{44}$ & 78 & 1.5$^{+2.2}_{-2.2}\times$10$^{45}$ & 39$^{+6}_{-6}$ & 1\\
  1.13 & 2.2$\times$10$^{43}$ & 4.1$\times$10$^{44}$ & 78 & 1.3$^{+0.2}_{-0.2}\times$10$^{45}$ & 34$^{+6}_{-6}$ & 1\\
  1.14 & 2.5$\times$10$^{43}$ & 4.9$\times$10$^{44}$ & 75 & 1.5$^{+0.4}_{-0.3}\times$10$^{45}$ & 40$^{+9}_{-9}$ & 0\\
  1.14 & 2.8$\times$10$^{43}$ & 5.8$\times$10$^{44}$ & 68 & 1.3$^{+0.2}_{-0.2}\times$10$^{45}$ & 35$^{+6}_{-6}$ &  0\\
  1.17 & 3.4$\times$10$^{43}$ & 7.3$\times$10$^{44}$ & 88 & 1.1$^{+0.2}_{-0.2}\times$10$^{45}$ & 29$^{+6}_{-6}$ & 1\\
  1.14 & 3.9$\times$10$^{43}$ & 8.7$\times$10$^{44}$ & 73 & 1.5$^{+0.3}_{-0.3}\times$10$^{45}$  & 40$^{+8}_{-8}$ & 0\\
  1.11 & 4.4$\times$10$^{43}$ & 1.0$\times$10$^{45}$ & 80 & 9.6$^{+1.6}_{-1.6}\times$10$^{44}$  & 25$^{+4}_{-4}$ & 0\\
  1.13 & 5.0$\times$10$^{43}$ & 1.2$\times$10$^{45}$ & 65 & 1.5$^{+0.2}_{-0.2}\times$10$^{45}$  & 39$^{+5}_{-5}$ & 0\\
  1.14 & 6.0$\times$10$^{43}$ & 1.6$\times$10$^{45}$ & 75 & 1.9$^{+0.5}_{-0.5}\times$10$^{45}$ & 50$^{+13}_{-13}$ & 0\\
  1.19 & 7.0$\times$10$^{43}$ & 2.0$\times$10$^{45}$ & 85 & 1.6$^{+0.2}_{-0.2}\times$10$^{45}$  & 41$^{+6}_{-6}$ & 1\\
  1.16 & 8.5$\times$10$^{43}$ & 2.6$\times$10$^{45}$ & 68 & 1.5$^{+0.3}_{-0.3}\times$10$^{45}$ & 39$^{+7}_{-7}$ & 1\\
  1.2 & 1.1$\times$10$^{44}$ & 3.9$\times$10$^{45}$ & 78 & 1.4$^{+0.3}_{-0.3}\times$10$^{45}$ & 37$^{+7}_{-7}$ & 0\\
  1.2 & 1.5$\times$10$^{44}$ & 5.9$\times$10$^{45}$ & 58 & 2.3$^{+0.4}_{-0.4}\times$10$^{45}$ & 60$^{+11}_{-11}$ & 0\\
  1.31 & 4.3$\times$10$^{44}$ & 4.5$\times$10$^{46}$ & 68 & 3.5$^{+0.7}_{-0.7}\times$10$^{45}$ & 91$^{+18}_{-18}$ & 0\\
  \hline 
  1.88 & 3.2$\times$10$^{42}$ & 4.0$\times$10$^{43}$ & 61 & 3.3$^{+0.6}_{-0.6}\times$10$^{45}$ & 88$^{+15}_{-16}$ & 0\\
  1.83 & 1.1$\times$10$^{43}$ & 1.7$\times$10$^{44}$ & 73 & 3.2$^{+0.6}_{-0.6}\times$10$^{45}$ & 86$^{+16}_{-16}$ & 0\\
  1.86 & 2.4$\times$10$^{43}$ & 4.6$\times$10$^{44}$ & 85 & 3.6$^{+0.6}_{-0.6}\times$10$^{45}$  & 94$^{+16}_{-16}$ & 1\\
  1.9 & 3.0$\times$10$^{43}$ & 6.3$\times$10$^{44}$ & 76 & 4.2$^{+0.6}_{-0.6}\times$10$^{45}$ & 112$^{+16}_{-16}$ & 0\\
  1.88 & 3.8$\times$10$^{43}$ & 8.5$\times$10$^{44}$ & 81 & 3.8$^{+0.8}_{-0.8}\times$10$^{45}$ & 101$^{+21}_{-21}$ & 0\\
  2.02 & 7.3$\times$10$^{43}$ & 2.1$\times$10$^{45}$ & 83 & 4.4$^{+0.6}_{-0.6}\times$10$^{45}$ & 116$^{+16}_{-16}$ & 1\\
  1.94 & 8.5$\times$10$^{43}$ & 2.6$\times$10$^{45}$ & 78 & 3.7$^{+0.5}_{-0.5}\times$10$^{45}$ & 98$^{+12}_{-12}$ & 0\\
  1.95 & 1.2$\times$10$^{44}$ & 4.4$\times$10$^{45}$ & 85 & 3.8$^{+0.9}_{-0.7}\times$10$^{45}$ & 100$^{+25}_{-19}$ & 0\\
  1.89 & 1.4$\times$10$^{44}$ & 5.7$\times$10$^{45}$ & 71 & 5.6$^{+1.1}_{-1.1}\times$10$^{45}$ & 148$^{+28}_{-29}$ & 0\\
  2.01 & 2.1$\times$10$^{44}$ & 1.0$\times$10$^{46}$ & 81 & 3.2$^{+0.6}_{-0.6}\times$10$^{45}$ & 86$^{+15}_{-15}$ & 0\\
  1.94 & 2.6$\times$10$^{44}$ & 1.4$\times$10$^{46}$ & 76 & 5.7$^{+0.6}_{-0.6}\times$10$^{45}$ & 150$^{+16}_{-16}$ & 0\\
  1.91 & 3.6$\times$10$^{44}$ & 2.5$\times$10$^{46}$ & 85 & 3.1$^{+0.5}_{-0.6}\times$10$^{45}$ & 82$^{+14}_{-15}$ & 0\\
  2.09 & 8.0$\times$10$^{44}$ & 1.2$\times$10$^{47}$ & 83 & 4.8$^{+1.2}_{-1.2}\times$10$^{45}$ & 86$^{+32}_{-33}$ & 1\\
\hline
\end{tabular}
        \caption{The average redshift, X-ray luminosity, AGN bolometric luminosity, IR
          luminosity due to star formation, and SFR, for the data presented in Figure \protect\ref{meanLirLagn}. 
          The errors on the $\mlir$ are calculated using the bootstrap analysis (see \S \protect\ref{meantech}). 	
          We also provide the censorship level of each bin, and a flag indicating when the minimum 
          value of the sources in that bin is an upper limit (when the flag has a value of 1), which can result to an extra uncertainty on the $\mlir$ (see \S\protect\ref{meantech}).}
          \label{tabresults} 
      \end{center}
    \end{table*}

\begin{figure}
			\begin{center}
				\includegraphics[scale=0.32]{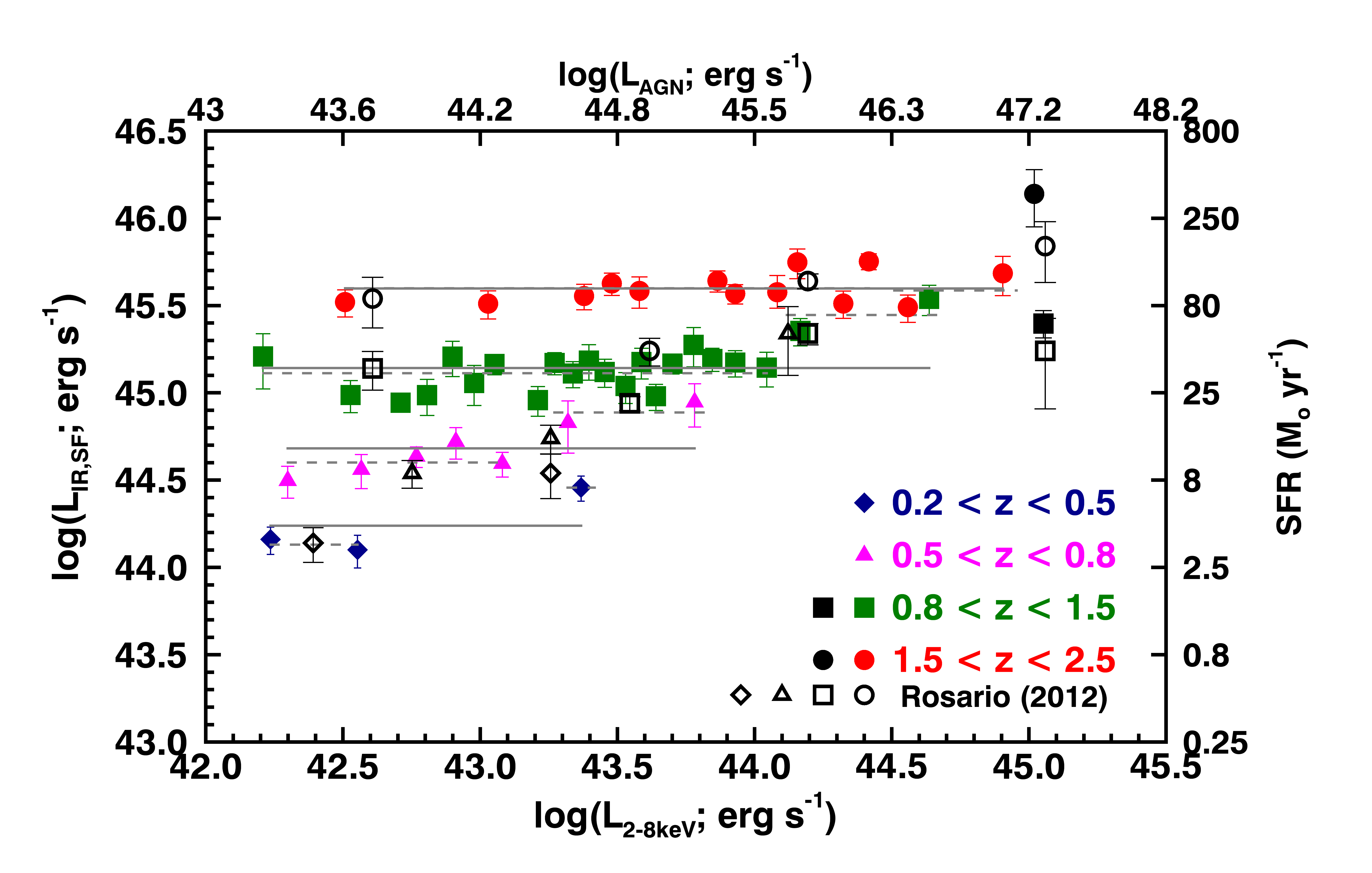}
 				\caption{$\mlir$ as a function of
                                  $\mlx$, as plotted in Figure
                                  \protect\ref{meanLirLagn} (also to be referred to for axis definitions). The horizontal 
                                  grey lines indicate the overall mean 
                                  $\lir$ across all of the $\lx$ bins 
                                  for each redshift range. The dashed grey 
                                  lines indicate the mean $\lir$ for (1) the one or two
                                  highest $\lx$ bins and (2) the lower $\lx$ bins 
                                  for each redshift range (see \S\protect\ref{meansfr_lx}). 
                                  The black hollow symbols are the stacking results of
                                   \protect\cite{Rosario12}, and the 
                                  black filled symbols are bins of the highest $\lx$ sources 
                                  from our study (we note that there are very few sources 
                                  in these bins for both studies; see \S\protect\ref{meansfr_lx}).
                                  Our results are broadly consistent with a flat
                                  relationship; however, for the redshift ranges with $z <$ 1.5 
                                  the highest $\lx$ bins are systematically a factor of 
                                  $\sim$2 higher than the mean $\lir$.
                                  }
				\label{Lirfitline}
			\end{center}
	\end{figure}

%Compare to previous studies
We find that our results are in general agreement to those studies that stack the 
FIR data to derive SFRs using large numbers of sources (e.g., \citealt{Harrison12}, \citealt{Rosario12});
however our results have reduced scatter and reduced uncertainties on the
AGN contribution to the IR luminosity. We look in more detail 
at how our results compare to those of stacking in the Appendix of this paper.
Additionally, we compare our results directly to those of \cite{Rosario12}, who explore the average 
60$\um$ luminosity ($\rm \nu L_{60\um}$) values (as a tracer of SFR) in the same redshift ranges as our study,
by stacking \textit{Herschel}--PACS data. 
We use the average difference between $\rm \nu L_{60\um}$ and  $\lir$ from our 
SED fitting results, $\lir/\rm \nu L_{60\um} = $ 2.2, to convert the results of \cite{Rosario12} to $\lir$. 
In Figure \ref{Lirfitline} we plot our results in comparison to those of \cite{Rosario12} 
(hollow black symbols) and find broad agreement with our results both as a function of redshift and 
$\lx$, although we have more $\lx$ bins and our results show less scatter. 
To compare to the highest $\lx$ bins of \cite{Rosario12} at the redshift ranges of 
$z =$ 0.8 -- 1.5 and $z =$ 1.5 -- 2.5
we calculate the $\mlir$ for the five highest $\lx$ sources in our study 
in both of these redshift ranges (plotted in Figure \ref{Lirfitline} with solid black symbols). We find that our 
highest $\lx$ sources are in agreement with those of \cite{Rosario12}; however, due to the 
very small number of sources in these bins (5 -- 23 sources across both studies), 
we do not interpret them any further.
  
%--> Assume all upper limits are 0 absolute values
To asses the contribution of the upper limits on the overall mean, 
we take an extreme scenario where all upper limits are 
assumed to correspond to zero values. We find that $\mlir$ can drop 
by 0.2\,dex (factor of 1.6) at 0.2 $< z <$ 0.5, by 0.3\,dex (factor of 2) at 0.5 $< z <$ 0.8 and 0.8 $< z <$ 1.5,
and by 0.4\,dex (factor of 2.5) at the highest redshift range of 1.5 $< z <$ 2.5.
However, we note that the form of the observed \textit{flat relationship} of $\mlir$ with $\lx$ 
(Figure \ref{meanLirLagn}) shows little to 
no change for all redshift ranges, in this extreme scenario. 
 
 %Compare to flat trend 
To test whether our results are consistent with a flat trend of $\mlir$ with
$\lx$ we show in Figure \ref{Lirfitline}, as a horizontal grey line, 
the mean $\mlir$ for each redshift range.
Across all redshifts the data lie within a factor of 2 of the mean. 
However, we find that the $\mlir$
values of the most luminous AGN for all of the redshift ranges at $z <$ 1.5 
are systematically above the overall mean. To quantify the 
deviation between the $\mlir$ of the high $\lx$ bins to the rest of
the data we make two fits; one to the two highest $\lx$ bins (with the
exception of $z =$ 0.2 -- 0.5 where we use only the highest $\lx$ bin); and 
one to the rest of the luminosity bins in the same redshift range 
(see the grey dashed lines of Figure \ref{Lirfitline}). We find an 
increase in $\mlir$ by a factor of $\sim$2 for the highest $\lx$ when compared to the lower
$\lx$ bins in each of the redshift ranges with $z <$ 1.5. For $z =$ 1.5 -- 2.5 there is no
significant difference in $\mlir$ between the highest and lowest $\lx$ that we cover.
We note that the systematic increase of $\mlir$ at high $\lx$ values observed in
the redshift ranges of $z <$ 1.5 does not correspond to a systematic
increase of the redshifts at high $\lx$ values (see Table \ref{tabresults}). Thus 
the modest trends observed at the high $\lx$ are not driven 
by redshift. We investigate the observed trends further in \S\ref{models}. 
 
 \subsection{Comparing to the average SFRs of the overall star forming galaxy population} \label{MS} 
\begin{figure}
			\begin{center}
				\includegraphics[scale=0.32]{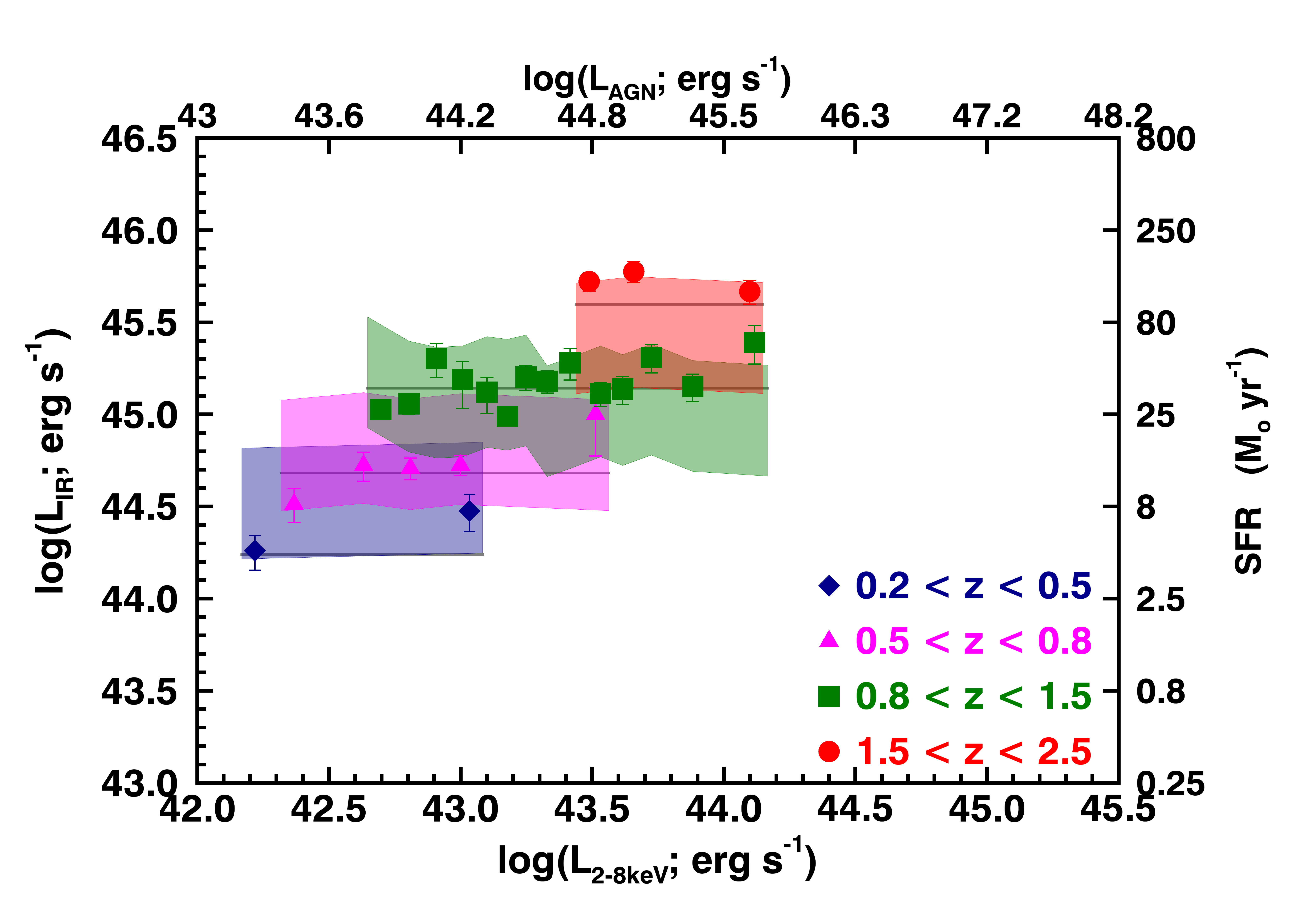}
 				\caption{$\mlir$ as a function of 
                                  $\mlx$ for the 
                                  subsample of sources that have
                                  a reliable stellar mass (M$_*$) measurement in
                                  \protect\cite{Ilbert13} (see \S\protect\ref{MS}; also see 
                                  Figure \protect\ref{meanLirLagn} for the axis definitions). 
                                  The grey solid lines are 
                                  the means for each redshift range of the whole parent 
                                  sample (see Figure \protect\ref{Lirfitline}). 
                                  The shaded regions
                                  correspond to the expected range in $\lir$
                                  for the overall star forming galaxy population
                                  at the mean redshift and mean M$_*$ of
                                  each bin
                                  as defined by \protect\cite{Schreiber14}. For all
                                  redshift ranges the $\mlir$ values of the
                                  AGN appear to be consistent with normal
                                  star forming galaxies.}
				\label{lirlxms}
			\end{center}
	\end{figure}

Here we explore whether X-ray AGN have SFRs that are consistent with being selected from the
overall star forming galaxy population. We compare the average SFRs of the AGN to the observed
relationship between SFR, redshift, and stellar mass (M$_*$) of normal 
star forming galaxies, which is defined
as the ``main sequence'' of star forming galaxies (e.g., \citealt{Noeske07};
\citealt{Elbaz11}; \citealt{Schreiber14}; \citealt{Speagle14}). To make this comparison we require 
stellar masses for the AGN in our sample. We use the
stellar masses from \cite{Ilbert13} for the sources in the
C-COSMOS area. Since their analysis did not take into account of a
possible AGN component to the rest--frame UV to near--IR SEDs, we applied a colour cut to
exclude sources for which there is likely to be significant AGN contamination to the SED.
 We only include AGN with rest frame colours $\rm U-V > 1$ and $\rm V-J > 1$ 
based on the analyses of \cite{Mullaney12b}. This results in a subsample of 
primarily moderate luminosity AGN ($\lx \lesssim$ 10$^{44} \ergss$) making up $\sim$40\% of the
parent sample at $z <$ 1.5, but only 26\% of the parent sample at $z
=$ 1.5 -- 2.5. For these sources, with a reliable M$_*$, 
we calculate the $\mlir$ as described in \S\ref{meantech}. 
Due to the reduced number of 
sources with masses we can no longer use bins of $\approx$40 
sources and we therefore reduce the number of
sources required in each bin to 25. We show the $\mlir$ as a function
of $\lx$, for the sub-sample with reliable M$_*$ values, in 
Figure \ref{lirlxms}. We note that this sub-sample have $\mlir$
values consistent with the whole parent sample (see Figure \ref{lirlxms}),
with the exception of the $z =$ 1.5 -- 2.5 range which appear to be
systematically higher.

We use the mean redshift and mean M$_*$ of each bin 
in Figure \ref{lirlxms} to calculate the expected range in
$\lir$ for ``main sequence'' galaxies using Equation 9 of \cite{Schreiber14}.
The shaded regions, colour-coded by redshift, correspond to the range of
$\lir$ covered by the main sequence galaxies at the mean redshift 
and mean M$_*$ of the sources in each bin; i.e. a scatter of 2 
around the mean results from \cite{Schreiber14}. 
We also find that these results are the same if we use the \cite{Elbaz11} 
definition of the ``main sequence''.
We find that, for this sample of X-ray AGN with $\lx \lesssim 10^{44} \ergss$, the $\mlir$ in all redshift
ranges with $z <$ 1.5 are consistent with that of star forming galaxies
of the same mean redshift and mass. This result agrees with the results of previous 
studies (e.g., \citealt{Mullaney12a}, \citealt{Harrison12}, \citealt{Bongiorno12}, 
\citealt{Lanzuisi15}). 
However, for the redshift range of $z =$ 1.5 -- 2.5 the $\mlir$ 
is systematically at the higher end of the $\lir$ region covered 
by ``main sequence'' galaxies, 
which may be due, in part, to a bias due to the fact
that only 26\% of the parent sample at those redshifts have reliable
masses, and these have systematically higher $\mlir$ values than the 
parent population (see Figure \ref{lirlxms}).   
 
\subsection{Comparing to empirical models} \label{models}
As shown in Figure \ref{Lirfitline}, the trend 
of $\mlir$ ($\langle$SFR$\rangle$) with $\lx$ 
($\lagn$) is broadly consistent with being flat. 
This result may initially seem in disagreement with the results of studies 
such as \cite{Rafferty11}, \cite{Mullaney12b},
\cite{ChenTing13}, \cite{Delvecchio14}, and \cite{Rodighiero15},
which find a correlation between the average $\lagn$ and 
SFR of star forming galaxies. However, these studies start with a 
parent population of \textit{galaxies} for which they calculate the average
 $\lagn$, while in this study we start with a population of \textit{AGN}
for which we calculate the average SFR.
It has been suggested that the variability of AGN, taking place on 
smaller timescales to that of star formation, could flatten any intrinsic 
correlation between the SFR and the $\lagn$ when not averaging over the 
most variable quantity (i.e. by taking the average  $\lagn$ over 
bins of SFR; e.g., \citealt{Hickox14}).
To assess what could be 
the driver of the flat relationship that we observe, and if indeed it is AGN variability that
is driving its shape, we compare to two empirical ``toy-models'' that predict the 
$\mlir$ as a function of $\lagn$. Firstly that of \cite{Hickox14} and secondly, a model 
based on \cite{Aird13} (also see \citealt{Caplar15} for a similar model).

	\begin{figure}
			\begin{center}
				\includegraphics[scale=0.32]{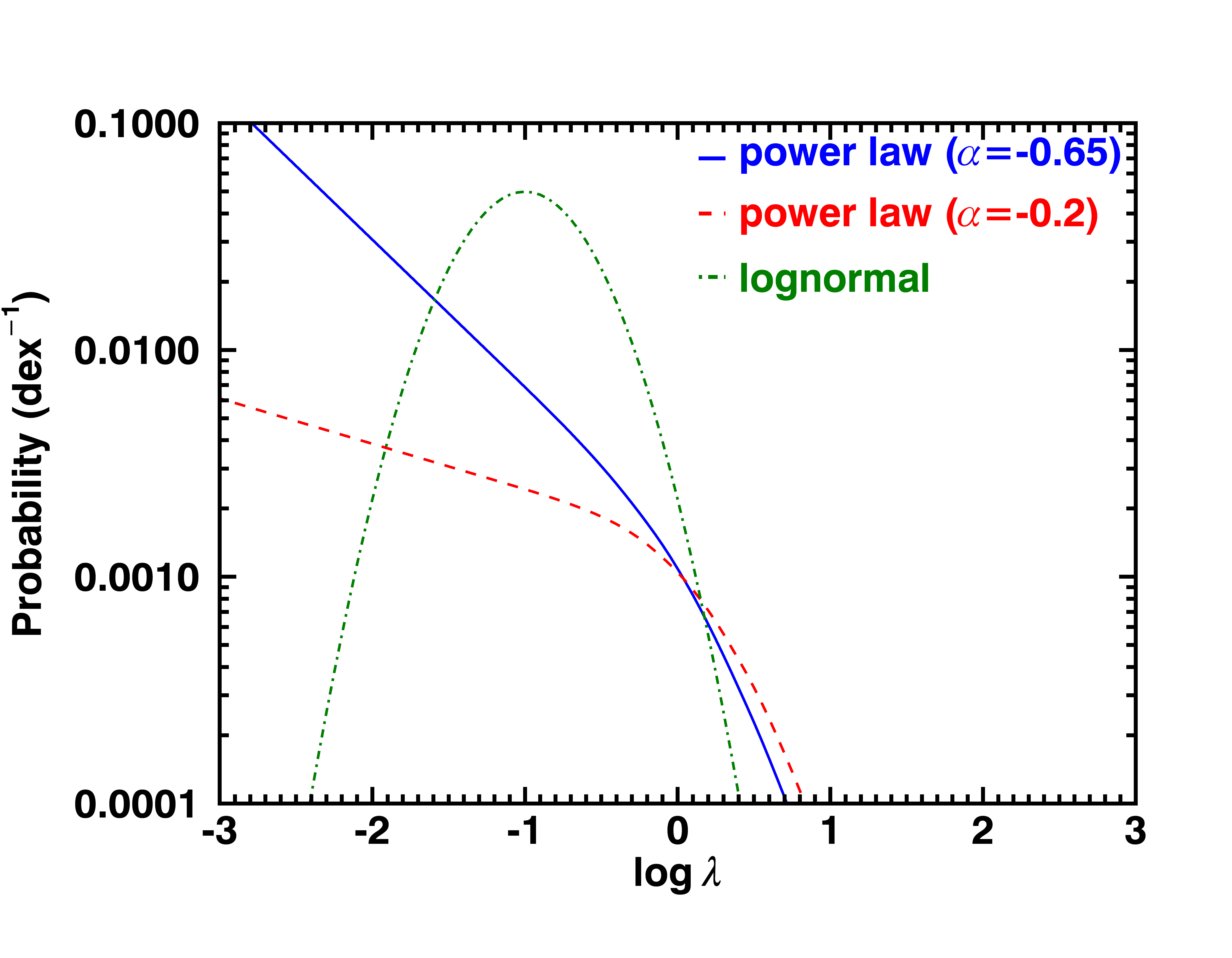}
 				\caption{
 				The probability distribution of the Eddington ratio ($\lambda$)
 				for the three cases assumed in Figure \protect\ref{comparingtomodels} 
 				(i.e. two broken power law distributions with a faint end slope of $\alpha = -$0.65 
 				and $\alpha = -$0.2, and a lognormal distribution with 0.4\,dex dispersion; see \S 2.2 in \protect\citealt{Aird13}).
 				This also serves as a \textit{schematic representation} 
 				of the three distributions assumed for the \protect\cite{Hickox14} model, assuming that the shape of the distributions
 				represent the variability function of individual AGN (see Section \ref{models} and Section 2 of \protect\citealt{Hickox14}).
 				}
				\label{eddratio}
			\end{center}
	\end{figure}

	\begin{figure}
		\begin{center}
		   \subfloat[]{\includegraphics[scale=0.32]{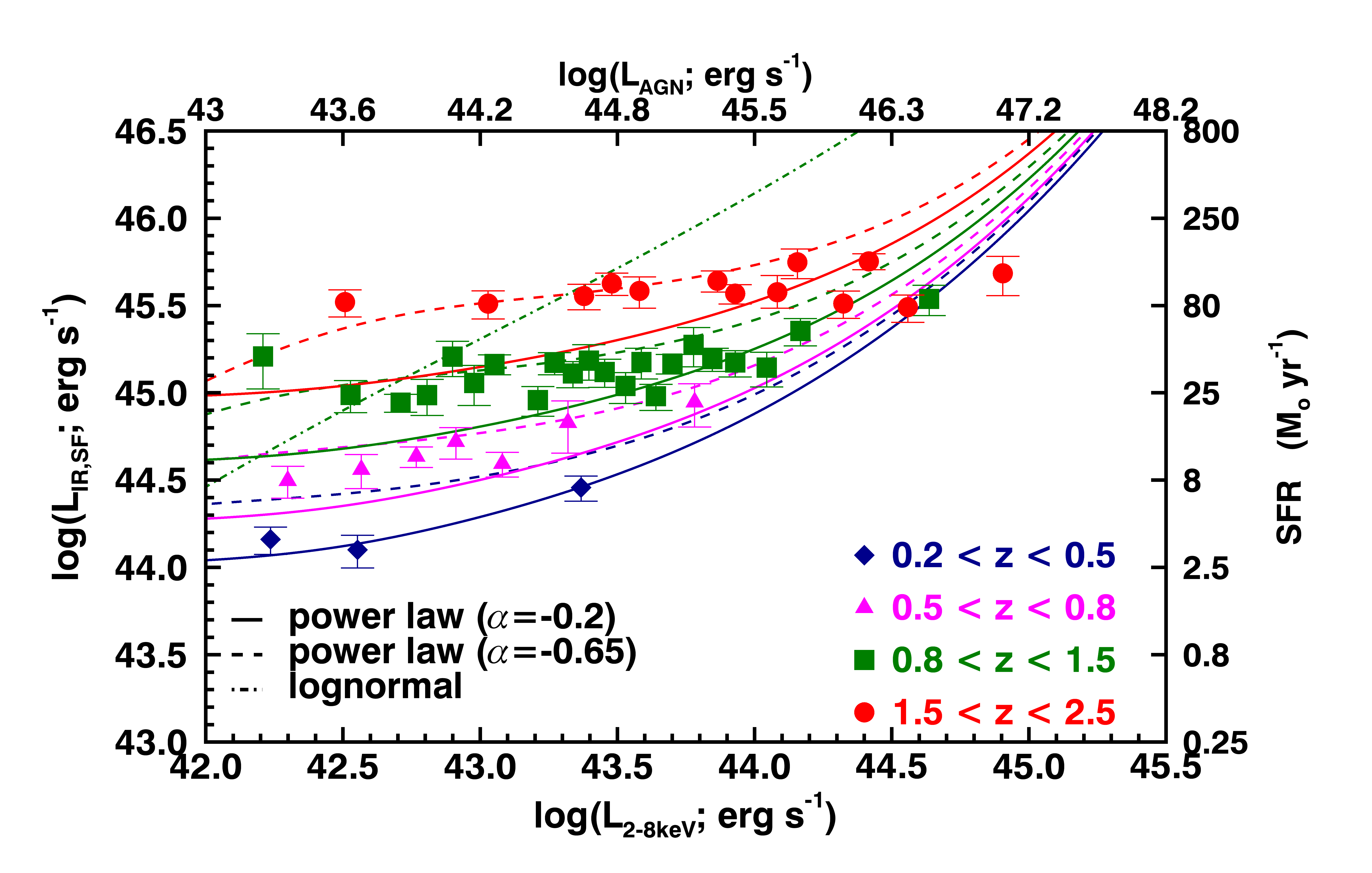}} \\
            \subfloat[]{\includegraphics[scale=0.32]{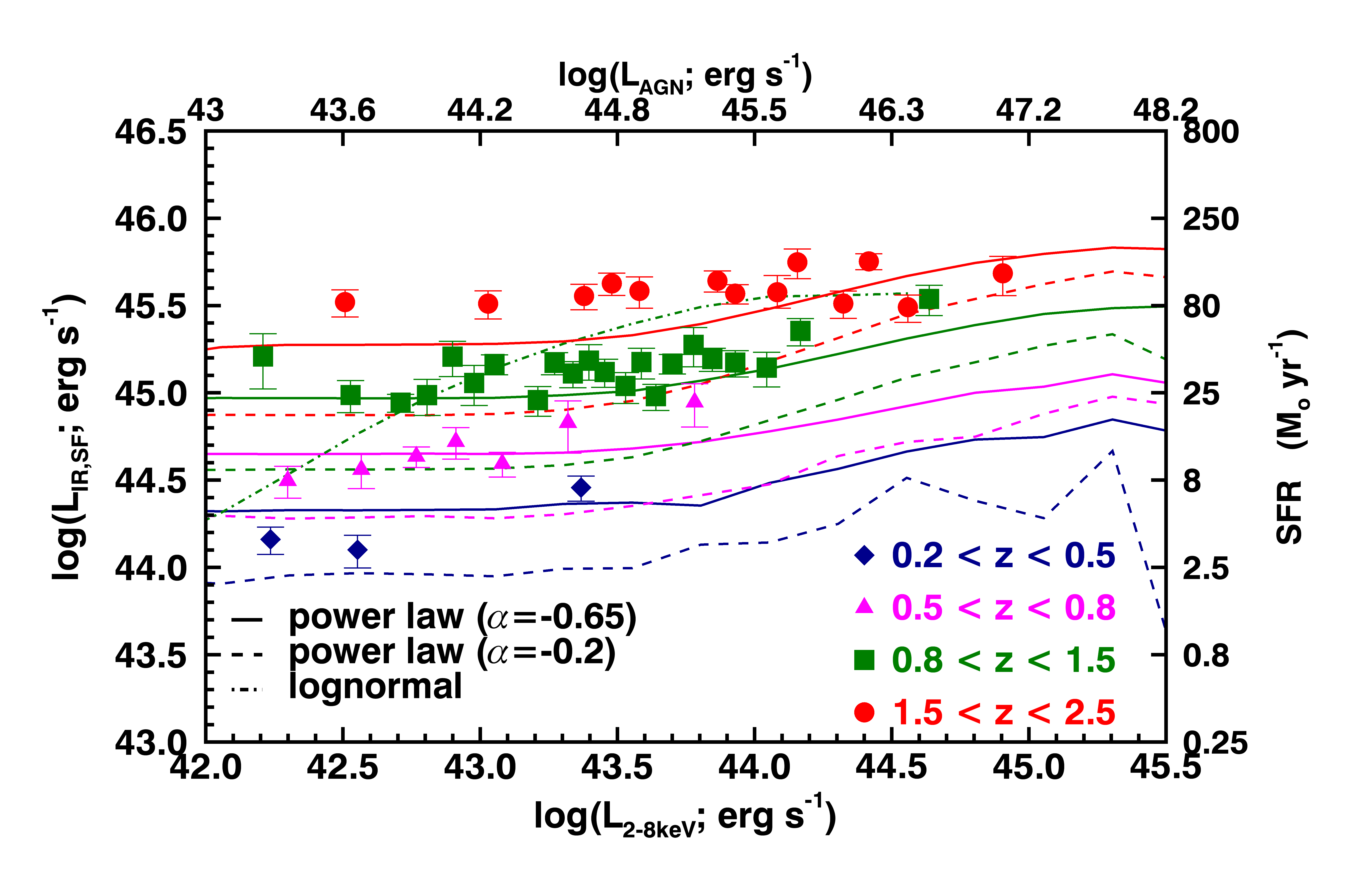}}  \\
			\caption{$\mlir$ as a function of $\mlx$ in four 
			redshift ranges compared to model tracks from (a) \protect\cite{Hickox14} and 
			(b) the extended \protect\cite{Aird13} model (see Figure \protect\ref{meanLirLagn} 
			for the axis definitions). The solid lines in both cases show the 
			predictions of the models with their originally assumed Eddington ratio distribution.
			From the two models the one of \protect\cite{Aird13}, which is based solely on observational 
			data, is in better agreement to our results; 
			however both models demonstrate how the flatness of the trends in our results are likely 
			to be a consequence of the assumed Eddington ratio distribution, or AGN variability. 
		    We also investigate how different the trends are when assuming different Eddington ratio distributions 
			in the two models (i.e. two broken power-law distributions with a faint end slope of 
			$\alpha = -$0.65 and $\alpha = -$0.2 respectively, and a lognormal distribution 
			for which we only show the tracks for 0.8$ < z < $1.5 to avoid confusion; see Figure \protect\ref{eddratio}). 
			The different assumed Eddington ratio distributions show significant differences in the predicted trends.
			See \S\protect\ref{models} for more details.}
				\label{comparingtomodels}
		\end{center}	
	\end{figure} 

The empirical ``toy-model'' presented in \cite{Hickox14}
assumes that SFR is correlated to $\lagn$
 when averaged over timescales of the 
order of 100\,Myr. To create the SFR distribution of the galaxies
in their model, they assume the redshift dependent
IR luminosity function (LF) from \cite{Gruppioni13}.
In the model, the individual AGN are allowed to
vary on short time scales on the basis of an assumed 
$\lagn /\langle \lagn \rangle$ distribution, 
which serves as a tracer of the Eddington ratio 
distribution of individual AGN
in the absence of black hole masses (see details in \citealt{Hickox14}). 
The fiducial model assumes that the distribution of
$\lagn/\langle \lagn \rangle$ has the form of a Schechter function 
(broken power law form) with a faint end slope of
$\alpha = -$0.2 and a cutoff luminosity of L$_{\rm cut}$\,=\,100\,$\langle
\lagn \rangle$ (see the dashed red curve in Figure \ref{eddratio} for a schematic of 
this distribution). The model can then predict the average SFR as a function
of instantaneous (i.e. observed) $\lagn$ of a large population of simulated AGN. 
We ran the model for the four redshift ranges of this study and
plot the results in Figure \ref{comparingtomodels}(a) with solid tracks. 
The model successfully reproduces an increase of the $\mlir$ with redshift, for a fixed range in $\lagn$,
\footnote{The increase of the $\mlir$ with redshift, for a fixed range in $\lagn$, could 
initially seem contradictory to the model's original assumption of a correlation of SFR 
and the long term averaged $\lagn$. However, even though the increase of $\mlir$ will 
be accompanied by an increase in the long term averaged $\lagn$, there is not
 a significant difference in the range of \textit{instantaneous} $\lagn$, across the 
simulated population, which is the quantity 
we effectively observe for an X-ray AGN sample.}
and is in good agreement with the data at $z =$ 0.2 -- 0.5; however, it fails to reproduce
the trends observed for the higher redshift ranges. In particular, the
normalisation of the predicted trends are too low compared to our data
and the rise of $\mlir$ with $\lagn$ is much steeper than that observed. 
The steepness of the predicted $\mlir$ trends at the highest $\lagn$
could be a result of the enforced correlation between SFR and
the long term $\mlagn$, or could be caused by the lack of an explicit 
Eddington limit in the model but rather a cut-off limit at high $\lagn$/SFR ratios
(see \citealt{Hickox14} for details).
We investigate how the predicted relationship varies with different 
variability prescriptions later on in this section.

The second empirical ``toy-model'' that we have compared to is based on \citet{Aird13},
which we extended to make predictions for the relationship
between AGN luminosity and star formation.
This model uses the observed redshift dependent
stellar mass function (SMF) of galaxies \citep[from
][]{Moustakas13} in combination with the probability function of a
galaxy of a given stellar mass and redshift hosting an AGN, based on
measurements in \cite{Aird12} for $ z  \lesssim  1$. This model predicts the
distribution of stellar masses, for which they correct to BH masses assuming
M$_{\rm BH} = 0.002\times$M$_*$ based on \cite{Marconi03}, as a function of X-ray luminosity.
In contrast to the \cite{Hickox14} model they use an Eddington
ratio distribution in the form of a broken power-law function with the faint 
end slope being steeper with $\alpha = -$0.65 (see the blue curve in Figure 
\ref{eddratio} for a schematic of this distribution).
\footnote{We note that \cite{Aird13} use an observed specific accretion rate 
distribution (i.e., $\lagn/$M$_*$) which they convert to an Eddington ratio distribution.}
In order to compare to our results we have extended
the model to cover the same redshift range as that of our sample and
convert the predictions of stellar mass to predictions of SFR. To
achieve this we adopt the measurements of the SMF by \cite{Ilbert13}
at $z =$ 1 -- 2.5 as an extension of the \cite{Moustakas13} SMF  
up to $z =$ 1, and extrapolate
the redshift-dependence of the probability of a galaxy hosting an AGN
from \cite{Aird12} to $z >$ 1 (which is consistent with the $z >$ 1 measurements
from \citealt{Bongiorno12}). Furthermore, we make the assumption that all
of the AGN are hosted by normal star forming
galaxies that lie on the ``main sequence'' as derived by \cite{Elbaz11},
which is motivated by the results of our study (see \S\ref{MS}).
\footnote{We note that there is evidence in optical studies of X-ray AGN, such as \cite{Azadi14}, 
that a small fraction of these AGN are hosted by non star forming galaxies; however, \cite{Azadi14} 
find that these AGN appear to form a minority of the population and therefore we do not 
expect them to significantly affect our mean SFRs.}
We convert from the model predicted stellar masses to SFRs, allowing for a scatter of
0.3 dex in SFR around the ``main sequence'' relation. In Figure
\ref{comparingtomodels}(b) we present the resulting predictions of
$\mlir$ as a function of $\lagn$, plotted with solid lines, in comparison to our results for each
of the four redshift ranges. The predicted trends of the mean SFR in this case are flat
for a wide range of $\lagn$, similar to our data, with a slight rise in
$\mlir$ at high $\lagn$ (i.e., $\lagn \, \gtrsim \, 10^{45} \, \ergss$). On the basis of this modified \cite{Aird13} model, the slight 
rise of $\mlir$ observed in our data (see \S\ref{meansfr_lx}) may be driven by a small increase in the average masses 
of the galaxies hosting very luminous AGN.

To first order, the data are better described by the extended
\cite{Aird13} model than the \cite{Hickox14} model; see the solid
tracks in Figure \ref{comparingtomodels}(b) compared to those in
Figure \ref{comparingtomodels}(a). However, since the two models have
assumed different Eddington ratio distributions (or, equivalently,
$\lagn /\langle \lagn \rangle$ for the \citealt{Hickox14} model) we
also explore how sensitive the results are to this assumption. We
therefore also ran the models with a series of three
different Eddington ratio distributions to understand how sensitive
the predicted trends of $\mlir$ with $\mlagn$ are on the assumed
Eddington ratio distribution.  We used (1) a broken power-law
with $\alpha = -$0.2 (i.e, the fiducial distribution assumed by
\citealt{Hickox14}); (2) a broken power-law with $\alpha = -$0.65
(i.e, the fiducial distribution assumed by \citealt{Aird13}); and (3)
a narrow lognormal distribution with a dispersion of $\sim$0.4\,dex
centred at an Eddington ratio of $\sim 0.06$, as defined by
\cite{Kauffmann09} for nearby AGN residing in star forming
galaxies. These three distributions can be seen in Figure
\ref{eddratio}.

In Figure \ref{comparingtomodels}(a)\&(b) we show the three sets of
tracks which correspond to the resulting trends of $\mlir$ with
$\mlagn$ for the different assumptions of the Eddington ratio
distributions. A clear correlation between $\mlir$ and $\mlagn$ is
predicted for the lognormal distribution while, by comparison, the
power-law models predict a much flatter relationship. With a change of
power-law slope from $\alpha = -$0.2 to $\alpha = -$0.65, the
normalisation of the model tracks increase and the trend becomes
flatter. The different shapes of the model tracks are driven by the
relative difference between the low Eddington ratio slope and the
slope of the low-mass end of the galaxy SMF (i.e.,\ for $M<M_*$,
$\alpha\sim$~0). The predicted correlation between $\mlir$ and
$\mlagn$ for the lognormal distribution is due to the narrow range of
probable Eddington ratios. For the assumptions behind our
models when assuming the lognormal distribution, most of the AGNs are
accreting at a broadly similar Eddington ratio and therefore an
increase in $\lagn$ is predominantly due to an increase in
stellar mass (and hence SFR since we assume the main sequence of
star-forming galaxies). By contrast, the steep low-Eddington ratio
slope for the power-law models, when compared to the low-mass end
slope of the galaxy SMF, allows for a broad range of Eddington ratios
across a narrow range in stellar mass; i.e.,\ there is a higher
probability for an AGN of a given luminosity to be hosted in a
high-mass galaxy with a low Eddington ratio than a low-mass galaxy
with a high Eddington ratio. Indeed, on the basis of the extended
\cite{Aird13} model, the population of low-to-moderate luminosity AGN
($\lagn \, \lesssim \, 10^{45} \ergss$) predominantly reside in
galaxies of similar stellar mass ($M_* \sim 10^{10.5-11} M_\odot$),
and thus similar SFRs, but with a wide range of possible Eddington
ratios.

Overall, our results suggest that the observed trends of
$\langle$SFR$\rangle$ -- $\lagn$ are due to AGN being highly variable
and residing, on average, in normal star forming galaxies. 
Similar results have also been found by
hydrodynamical simulations that show that AGN variability can cause a
flat trend between $\lagn$ and SFR (e.g., \citealt{Gabor13};
\citealt{Volonteri15}). The
Eddington ratio distributions of AGNs are typically constructed to
describe a population of AGN. However, as adopted in our models, they
can also be understood as the distribution of Eddington ratios for
an individual AGN over time, and hence could be used as a variability
prescription of the AGN (as originally adopted in
\citealt{Hickox14}). As is clear from Figure
\ref{comparingtomodels}(a)\&(b), the choice of Eddington ratio
distribution plays a major role in the form of the predicted $\langle$SFR$\rangle$
-- $\lagn$ relationship. For example, our results are much better described
with the use of a broken power-law Eddington ratio distribution with a
faint end slope of $\alpha = -$0.65, than with a narrow lognormal
Eddington ratio distribution, which predicts a qualitatively different
$\langle$SFR$\rangle$ -- $\lagn$ relationship to that found from our
data. Thus, the $\langle$SFR$\rangle$ (or $\mlir$) -- $\lagn$ plane can
be a useful diagnostic tool for placing constraints on the intrinsic
Eddington ratio distribution of AGN (also see \citealt{Veale14}).

\section{Conclusions}
We have created a large sample of X-ray detected AGN
with FIR coverage and individual SFR measurements.
Our sample has a total of 2139 AGN at redshifts of $z =$ 0.2--2.5, with
$10^{42} < \lx < 10^{45.5} \ergss$. 
Using the available photometry from 8--500$\um$ we have performed
individual SED fitting to all of the sources in our sample,
and measure the IR luminosity due to star formation, $\lir$.  

Our analysis has a number of key advantages over many previous studies:
(a) the use of deblended source catalogues for the FIR
photometry, which ensures better constraints on the
flux density measurements and eliminates the overestimation due to blending and
confusion of sources (see \S\ref{mfir});
(b) the use of photometric upper limits in the SED fitting
analysis, which achieve better constraints on the fitted SEDs (see \S\ref{sed_an});
(c) the decomposition of the AGN and star formation contributions to the FIR emission,
which provides values of $\lir$ that are not contaminated by the
AGN (see \S\ref{sed_an});          
(d) the calculation of upper limits on $\lir$ when
the data are insufficient to identify the star forming component directly 
(i.e., not enough photometric data points, poor S/N data,
or dominant AGN component), which allows  
us to estimate the $\mlir$ for all the sources in our sample
avoiding the bias that could be caused by removing these sources (see \S\ref{meantech}).

With the $\lir$ measurements for each source we derived 
the mean $\lir$ values ($\mlir$; a proxy of the $\langle$SFR$\rangle$) as a function of 
$\lx$ (a proxy of the AGN luminosity; $\lagn$) in bins of $\sim$40 sources, for the redshift ranges 
of 0.2 -- 0.5, 0.5 -- 0.8, 0.8 -- 1.5, and 1.5 -- 2.5. In comparison to previous studies, our results show 
less scatter amongst $\mlir$ across the wide range of $\lx$ investigated in this study.
Overall we found that: 

\begin{enumerate}
	\item The $\langle$SFR$\rangle$ increases by more than an order of magnitude
          from redshifts of 0.2 -- 0.5 to 1.5 -- 2.5, in agreement with
          previous studies on the redshift evolution of the SFR for the general star forming
          galaxy population. See \S\ref{meansfr_lx}.
	\item For each redshift range the $\langle$SFR$\rangle$ shows no strong
          dependence on AGN luminosity; however we note that for the redshift
          ranges of $z \leq$ 1.5 the highest $\lagn$ systems have
          $\langle$SFR$\rangle$ values that are systematically higher than those of  
          lower $\lagn$ systems by a factor of $\approx$2. See \S\ref{meansfr_lx}.
	\item For the $\sim$40\% of the sources within the COSMOS area
          with reliable stellar masses, we
          compare their $\langle$SFR$\rangle$ to the ``Main Sequence" 
          of the overall star forming galaxy population.
          The X-ray AGN, at all redshift ranges, have $\langle$SFR$\rangle$ that are consistent with
          normal star forming galaxies at the same redshifts and masses. 
	      Due to a lack of secure masses for the high $\lagn$ systems in
          our sample this result is restricted to moderate AGN luminosities 
          (i.e., $\lx \lesssim$ 10$^{44.2} \ergss$ or $\lagn \lesssim$ 10$^{45.5} \ergss$). See \S\ref{MS}.
	\item To qualitatively understand the flat relationship
	    between the $\langle$SFR$\rangle$ and $\lagn$
	    we compared to two empirical ``toy-models''
        that make predictions for this relation: \cite{Hickox14} and an
        extended version of \cite{Aird13}. These models take mock galaxy populations
        and assign them with SFR values based on observed distributions, and instantaneous 
        $\lagn$ values based on an assumed Eddington ratio distribution.
        We find that the flat relationship seen in our data
       could be due to short timescale variations 
        in the mass accretion rates, which, in combination with the relative 
        shapes of the Eddington ratio distribution and the galaxy SMF, can wash 
        out the long term relationship between $\langle$SFR$\rangle$ and $\lagn$. See \S\ref{models}
	\item We find that the predicted $\langle$SFR$\rangle$ -- $\lagn$ relationship is sensitive to the assumed 
	    Eddington ratio distribution. 
	    For example, both models predict a relatively 
        flat relationship over all redshift ranges, assuming an Eddington ratio distribution
        of a broken power-law form with a faint end slope
        of $\alpha = -$0.65, whilst with a log-normal distribution the predicted trends 
        are too steep to be consistent with our data.  Therefore, the observed 
	    $\langle$SFR$\rangle$ -- $\lagn$ relationship appears to be a sensitive diagnostic 
	    of the intrinsic Eddington ratio distribution of AGN. 
	    See \S\ref{models}.
\end{enumerate} 

%%%%%%%%%%%%%%%%%%%%%%%%%%%%%%%%%%%%%%%%%%%%%%%%%
\subsection*{ACKNOWLEDGMENTS}
We thank the anonymous referee for their constructive comments
on the paper. We acknowledge the Faculty of Science Durham Doctoral Scholarship (FS),
the Science and Technology Facilities Council (CMH, DMA, AMS and ADM
through grant code ST/I001573/1), and the Leverhulme Trust (DMA). 
JAA acknowledges support 
from a COFUND Junior Research Fellowship from the Institute of 
Advanced Study, Durham University.
JRM acknowledges support from the University of Sheffield via its 
Vice-Chancellor Fellowship scheme.

%%%%%%%%%%%%%%%%%%%%%%%%%%%%%%%%%%%%%%%%%%%%%%%%%

%%%%%%%%%%%%%%%%%%%%%%%%%%%%%%%%%%%%%%%%%%%%%%%%%

\appendix
\section{Comparison of the K-M method to the stacking analysis method}
In this Appendix we compare our results using our SED fitting analysis and 
the K-M method that weused in this work (see Section \ref{analysis}), to 
those we would obtain with stacking analysis, a method commonly
used in similar studies of star-forming and AGN galaxy samples. 
 
%-->Stacking comparison
Following the method of \cite{Harrison12} we stacked the SPIRE-250$\um$ 
maps at the X-ray positions of the sources of our sample in C-COSMOS.
We use the C-COSMOS sample since it makes up most of our overall sample and avoids
issues that can arise when combining stacks of different fields with different depths.

\begin{figure}
		\begin{center}
                  \includegraphics[scale=0.35]{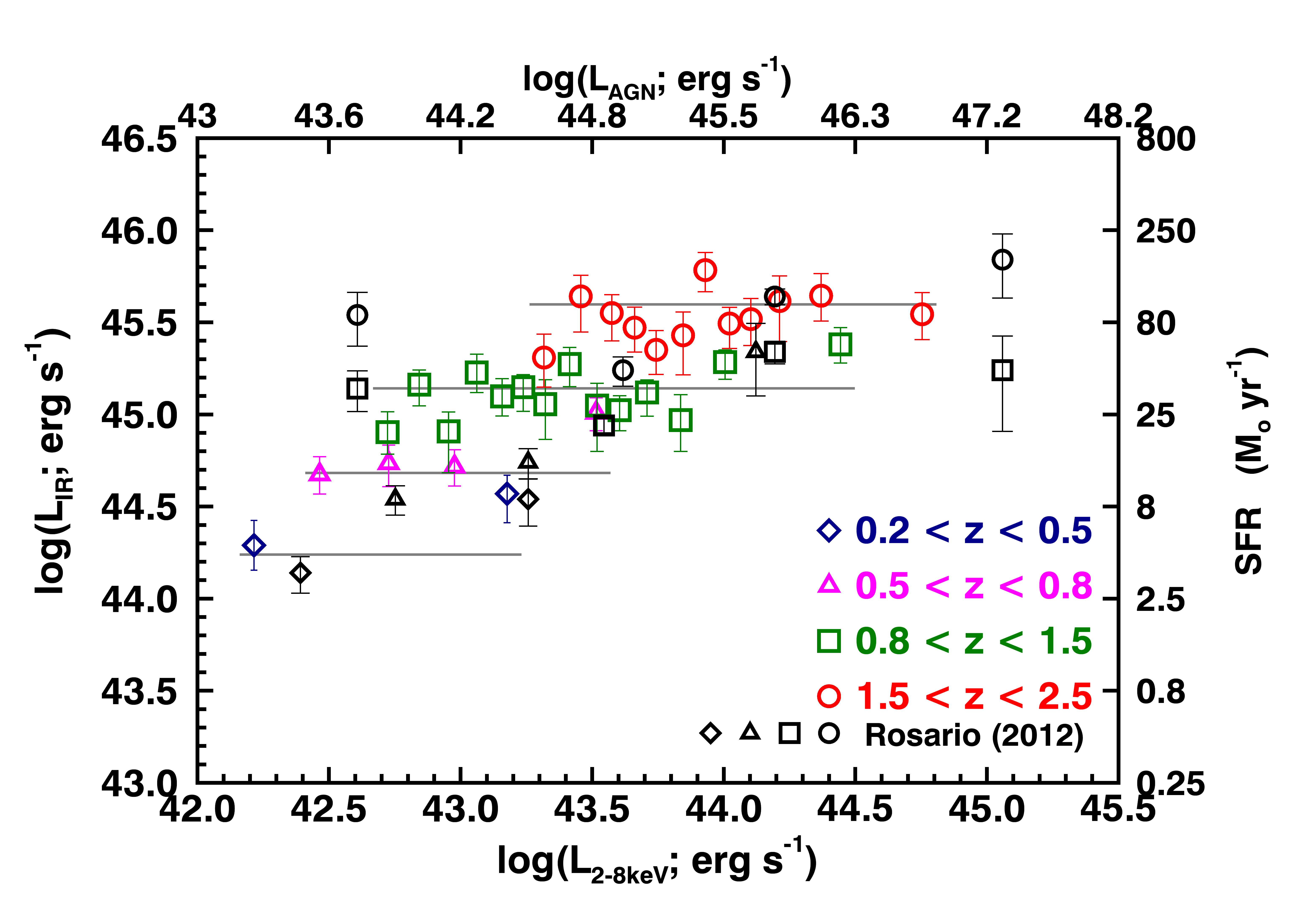}
                  \caption{$\mlir$ as a function of $\mlx$ when stacking the SPIRE data at 250$\um$ 
                  for the sources of C-COSMOS see 
                  Figure \protect\ref{meanLirLagn} for the axis definitions.
                  We compare these results to the overall K-M means of our SED results
                  (grey lines; see \S\protect\ref{meansfr_lx}), and the results of \protect\cite{Rosario12}.
                  We find that our results are consistent to those obtained using the stacking analysis, however 
                  the K-M method's results produce less scatter (see Figure \protect\ref{Lirfitline}).}
                  \label{stack}
		\end{center}	
\end{figure}

We bin the sample in bins of $\lx$ and reshift containing $\sim$40 sources each, in the
same way as described in \S \ref{meantech} for the K-M method (in the redshift range of $z =$ 0.2 -- 0.5 we use
$\sim$30 sources to allow for more than one bin). We show the stacking results in 
Figure \ref{stack}, in comparison with the overall means of the K-M method results for each redshift range,
as well as the results of \cite{Rosario12}. We find that our main results are consistent with 
the results we obtain when using the stacking analysis, and that both methods are in agreement with
the results of \cite{Rosario12}.

This comparison demonstrates that our method for calculating the mean produces results consistent to the popular 
method of stacking in the FIR. However, our method produces less scatter amongst bins,
as well as smaller uncertainties on the mean values. This is likely due to the use of deblended FIR 
photometry, and the removal of AGN contamination, in our analysis, which are effects not taken into when stacking.

\end{document}